\documentclass[notitlepage,preprintnumbers,prd,nofootinbib]{revtex4-1}

\usepackage{amsmath,amssymb,slashed}
\usepackage{hyperref}
\usepackage{url}
\usepackage{breakurl}
\usepackage{graphicx}
\usepackage{subfigure}

\begin{document}

\title{Hidden Pion Varieties in Composite Models for Diphoton Resonances}

\author{Keisuke Harigaya and Yasunori Nomura}
\affiliation{Berkeley Center for Theoretical Physics, Department of Physics, 
  University of California, Berkeley, CA 94720}
\affiliation{Theoretical Physics Group, Lawrence Berkeley National 
  Laboratory, Berkeley, CA 94720}

\begin{abstract}
The diphoton excesses at $750~{\rm GeV}$ seen in the LHC data 
may be the first hint for new physics at the TeV scale.  We discuss 
variations of the model considered earlier, in which one or more 
diphoton excesses arise from composite pseudo~Nambu-Goldstone bosons 
(hidden pions) associated with new strong dynamics at the TeV scale. 
We study the case in which the $750~{\rm GeV}$ excess arises from a 
unique hidden pion leading to a diphoton final state as well as the 
case in which it arises from one of the hidden pions decaying into 
diphotons.  We consider $SU(N)$, $SO(N)$, and $Sp(N)$ gauge groups 
for the strong dynamics and find that $SO(N)$ and $Sp(N)$ models give 
extra hidden pions beyond those in the $SU(N)$ models, which can be 
used to discriminate among models.
\end{abstract}

\maketitle

\section{Introduction}
\label{sec:intro}

The recently announced diphoton excess at $\simeq 750~{\rm 
GeV}$~\cite{ATLAS,CMS,ATLAS-2,CMS-2} may be the first hint 
for physics beyond the standard model at the TeV scale.  In 
Ref.~\cite{Harigaya:2015ezk}, we have proposed, based on stability 
of the theory and the strength of the signal, that this excess results 
from a composite spin-$0$ particle decaying into a two-photon final 
state.  In Ref.~\cite{Harigaya:2016pnu}, we have studied a particularly 
simple version of this, in which the $750~{\rm GeV}$ particle 
is a composite pseudo~Nambu-Goldstone boson associated with 
new strong dynamics at the TeV scale which is singly produced 
by gluon fusion and decays into two photons.  In particular, we 
have studied a model that has an extra gauge group $G_H = SU(N)$ 
at the TeV scale, in addition to the standard model gauge 
group $G_{\rm SM} = SU(3)_C \times SU(2)_L \times U(1)_Y$, 
with extra matter---hidden quarks---in the vectorlike bifundamental 
representation of $G_H$ and $SU(5) \supset G_{\rm SM}$.  We 
have studied detailed phenomenology of pseudo~Nambu-Goldstone 
bosons---hidden pions---one of which is the $750~{\rm GeV}$ diphoton 
resonance.  For related work involving similar dynamics, see 
Refs.~\cite{Nakai:2015ptz,Low:2015qep,Kilic:2009mi,Chiang:2015lqa,Redi:2016kip}.

In this paper, we study two classes of variations of the minimal model 
in Refs.~\cite{Harigaya:2015ezk,Harigaya:2016pnu}.  For definiteness, 
we keep the $G_{\rm SM}$ quantum numbers of the hidden quarks to be 
${\bf 5} + {\bf 5}^*$ of $SU(5) \supset G_{\rm SM}$, motivated by 
grand unification.  (Note that $SU(5)$ here is used as a mnemonic; 
it does not mean that the three factors of $G_{\rm SM}$ are 
actually unified at the TeV scale.)  In the first class, we consider 
models in which a single diphoton resonance arises from a unique 
$G_{\rm SM}$-singlet hidden pion.  After reviewing the $G_H = SU(N)$ 
model in Refs.~\cite{Harigaya:2015ezk,Harigaya:2016pnu}, we discuss 
models with $G_H = SO(N)$ and $Sp(N)$.  Other than being possible 
variations, these models have an additional motivation that the matter 
content is consistent with simple $SO(10)$ grand unified theories. 
We study symmetry breaking patterns and hidden pion contents, and 
we predict the spectra of hidden pions under the condition that the 
$G_{\rm SM}$-singlet hidden pion is responsible for the $750~{\rm GeV}$ 
excess.  We find that $SO(N)$ and $Sp(N)$ models have extra hidden 
pions beyond those in the $SU(N)$ model, which can be used to 
discriminate among models.  In fact, in terms of the hidden pion 
phenomenology, the models discussed here essentially cover the whole 
possibilities with the hidden quarks in a single representation of 
$G_H$ and ${\bf 5} + {\bf 5}^*$ of $SU(5) \supset G_{\rm SM}$.  In 
the second class, we consider models in which multiple (two) diphoton 
resonances arise from hidden pions.  In Ref.~\cite{Harigaya:2016pnu}, 
it was found that this can occur if the model contains an extra hidden 
quark that is charged under $G_H$ but singlet under $G_{\rm SM}$. 
(Introduction of such an extra hidden quark was motivated by cosmology 
there.)  We study $G_H = SU(N)$, $SO(N)$, and $Sp(N)$ models in this 
class.  We demonstrate how parameters of the models are determined 
and how other, $G_{\rm SM}$-charged hidden pion masses are predicted 
once the two diphoton resonances are observed.

The composite models we discuss contain would-be stable particles 
which do not decay solely by $G_H$ or $G_{\rm SM}$ gauge interactions.
If they are electrically or color charged, their lifetimes must be 
short enough to evade cosmological constraints.  The cosmological 
constraints and decays of would-be stable particles are discussed 
in Ref.~\cite{Harigaya:2016pnu} for $G_H=SU(N)$.  We extend these 
analyses to the case of $SO(N)$ and $Sp(N)$.

The organization of this paper is as follows.  In Section~\ref{sec:minimal}, 
we discuss models in which a single diphoton resonance arises from hidden 
pions.  Models with $G_H = SU(N)$, $SO(N)$, and $Sp(N)$ are considered 
in three subsections.  In Section~\ref{sec:non-minimal}, we discuss 
models in which two diphoton resonances arise from hidden pions. 
Again, models with $G_H = SU(N)$, $SO(N)$, and $Sp(N)$ are considered. 
In Section~\ref{sec:decays}, we discuss cosmological constraints and 
decays of the would-be stable particles.

\section{Minimal Models for the 750 GeV Resonance}
\label{sec:minimal}

Here we discuss variations of the minimal model in 
Refs.~\cite{Harigaya:2015ezk,Harigaya:2016pnu}.  After reviewing 
salient features of the model in Section~\ref{subsec:SU}, we present 
variations in which the hidden gauge group is changed from $SU(N)$ to 
$SO(N)$ and $Sp(N)$ in Sections~\ref{subsec:SO} and \ref{subsec:Sp}, 
respectively.  We discuss the symmetry breaking pattern and hidden 
pion spectrum in each case.  We find that the $SO(N)$ and $Sp(N)$ 
models have extra hidden pions beyond those in the $SU(N)$ case.

\subsection{{\boldmath $G_H = SU(N)$}}
\label{subsec:SU}

The model has a hidden gauge group $G_H = SU(N)$, with dynamical 
scale $\Lambda \approx O({\rm TeV})$, and hidden quarks charged under 
both $G_H$ and the standard model gauge groups, $G_{\rm SM}$, as in 
Table~\ref{tab:SU}.%
\footnote{Throughout the paper, we adopt the hypercharge normalization 
 such that the standard model left-handed Weyl fermions have 
 $(q,u,d,l,e) = (1/6,-2/3,1/3,-1/2,1)$.}
Here, we assume $N \geq 3$; the case of $G_H = SU(2)$ ($\simeq Sp(2)$) 
is analyzed in Section~\ref{subsec:Sp}.  The hidden quarks have mass terms
\begin{equation}
  {\cal L} = -m_D \Psi_D \bar{\Psi}_D - m_L \Psi_L \bar{\Psi}_L + {\rm h.c.},
\label{eq:L_mass}
\end{equation}
where we take $m_{D,L} > 0$ without loss of generality, and we assume 
$m_{D,L} \lesssim \Lambda$.  Note that the charge assignment of the 
hidden quarks is such that they are a vectorlike fermion in the 
bifundamental representation of $G_H$ and $SU(5) \supset G_{\rm SM}$, 
so that the model preserves gauge coupling unification at the level 
of the standard model.  Throughout the paper, we assume that the hidden 
sector preserves $CP$ to a good accuracy.
\begin{table}[h]
\begin{center}
\begin{tabular}{c|cccc}
   & $G_H = SU(N)$ & $SU(3)_C$ & $SU(2)_L$ & $U(1)_Y$ \\ \hline
 $\Psi_D$       &       $\Box$ &  ${\bf 3}^*$ & ${\bf 1}$ &  $1/3$ \\
 $\Psi_L$       &       $\Box$ &    ${\bf 1}$ & ${\bf 2}$ & $-1/2$ \\
 $\bar{\Psi}_D$ & $\bar{\Box}$ &    ${\bf 3}$ & ${\bf 1}$ & $-1/3$ \\
 $\bar{\Psi}_L$ & $\bar{\Box}$ &    ${\bf 1}$ & ${\bf 2}$ &  $1/2$ 
\end{tabular}
\end{center}
\caption{Charge assignment of the minimal model with $G_H = SU(N)$. 
 Here, $\Psi_{D,L}$ and $\bar{\Psi}_{D,L}$ are left-handed Weyl spinors. 
 We denote representations of $G_H$ by Young tableaux while those 
 of $G_{\rm SM}$ by the dimensions of representations.}
\label{tab:SU}
\end{table}

The strong $G_H$ dynamics makes the hidden quarks condensate
\begin{equation}
  \langle \Psi_D \bar{\Psi}_D + \Psi_D^\dagger \bar{\Psi}_D^\dagger \rangle 
  \approx \langle \Psi_L \bar{\Psi}_L 
    + \Psi_L^\dagger \bar{\Psi}_L^\dagger \rangle 
  \equiv - c.
\label{eq:PsiPsi-cond}
\end{equation}
These condensations do not break the standard model gauge groups, but 
they break the approximate $SU(5) \times SU(5)$ flavor symmetry of the 
$G_H$ gauge theory to the diagonal $SU(5)$ subgroup.  The spectrum below 
$\Lambda$ therefore consists of 24 hidden pions, whose quantum numbers 
under $G_{\rm SM} = SU(3)_C \times SU(2)_L \times U(1)_Y$ are
\begin{equation}
  \psi({\bf 8}, {\bf 1})_0, \qquad
  \chi({\bf 3}, {\bf 2})_{-5/6}, \qquad
  \varphi({\bf 1}, {\bf 3})_0, \qquad
  \phi({\bf 1}, {\bf 1})_0,
\label{eq:pions_SU}
\end{equation}
where $\psi$, $\varphi$, and $\phi$ are real scalars while $\chi$ is a 
complex scalar.  The masses of these particles are given by
\begin{align}
  m_\psi^2 &= 2 m_D \frac{c}{f^2} 
    + 3 \Delta_C,
\label{eq:m_psi}\\
  m_\chi^2 &= (m_D + m_L) \frac{c}{f^2} 
    + \frac{4}{3} \Delta_C + \frac{3}{4} \Delta_L + \frac{5}{12} \Delta_Y,
\label{eq:m_chi}\\
  m_\varphi^2 &= 2 m_L \frac{c}{f^2} 
    + 2 \Delta_L,
\label{eq:m_varphi}\\
  m_\phi^2 &= \frac{4 m_D + 6 m_L}{5} \frac{c}{f^2}.
\label{eq:m_phi}
\end{align}
Here, $f$ is the decay constant,%
\footnote{Our definition of the decay constant, $f$, is a factor of $2$ 
 different from that in Ref.~\cite{Weinberg:1996kr}:\ $f = F/2$.}
and $\Delta_{C,L,Y}$ are contributions from standard model gauge loops, 
of order
\begin{equation}
  \Delta_C \simeq \frac{3 g_3^2}{16\pi^2} \Lambda^2,
\qquad
  \Delta_L \simeq \frac{3 g_2^2}{16\pi^2} \Lambda^2,
\qquad
  \Delta_Y \simeq \frac{3 g_1^2}{16\pi^2} \Lambda^2,
\label{eq:Delta_CLY}
\end{equation}
where $g_3$, $g_2$, and $g_1$ are the gauge couplings of $SU(3)_C$, 
$SU(2)_L$, and $U(1)_Y$, respectively, with $g_1$ in the $SU(5)$ 
normalization.  Using naive dimensional analysis~\cite{Manohar:1983md}, 
we can estimate the quark bilinear condensate and the decay constant as
\begin{equation}
  c \approx \frac{N}{16\pi^2} \Lambda^3,
\qquad
  f \approx \frac{\sqrt{N}}{4\pi} \Lambda.
\label{eq:scales}
\end{equation}

The couplings of the hidden pions with the standard model gauge fields 
are determined by chiral anomalies and given by
\begin{align}
  {\cal L} =& \frac{N g_3^2}{64\pi^2 f} d^{abc} \psi^a 
    \epsilon^{\mu\nu\rho\sigma} G^b_{\mu\nu} G^c_{\rho\sigma} 
   + \frac{N g_3 g_1}{16\sqrt{15}\pi^2 f} \psi^a 
    \epsilon^{\mu\nu\rho\sigma} G^a_{\mu\nu} B_{\rho\sigma} 
\nonumber\\
  & - \frac{3 N g_2 g_1}{32\sqrt{15}\pi^2 f} \varphi^\alpha 
    \epsilon^{\mu\nu\rho\sigma} W^\alpha_{\mu\nu} B_{\rho\sigma} 
\nonumber\\
  & + \frac{N g_3^2}{32\sqrt{15}\pi^2 f} \phi\, 
    \epsilon^{\mu\nu\rho\sigma} G^a_{\mu\nu} G^a_{\rho\sigma} 
  - \frac{3 N g_2^2}{64\sqrt{15}\pi^2 f} \phi\, 
    \epsilon^{\mu\nu\rho\sigma} W^\alpha_{\mu\nu} W^\alpha_{\rho\sigma} 
  - \frac{N g_1^2}{64\sqrt{15}\pi^2 f} \phi\, 
    \epsilon^{\mu\nu\rho\sigma} B_{\mu\nu} B_{\rho\sigma},
\label{eq:pion-couplings}
\end{align}
where $a,b,c = 1,\cdots,8$ and $\alpha=1,2,3$ are $SU(3)_C$ and $SU(2)_L$ 
adjoint indices, respectively, and $d^{abc} \equiv 2 {\rm tr}[t^a \{ t^b, 
t^c \}]$ with $t^a$ being half of the Gell-Mann matrices.  Assuming that 
the $\phi$ particle produced by gluon fusion and decaying to a diphoton 
is responsible for the $750~{\rm GeV}$ excess, we find that the diphoton 
rate and $m_\phi \simeq 750~{\rm GeV}$ determine parameters of the model 
as~\cite{Harigaya:2016pnu}
\begin{align}
  f &\simeq 690~{\rm GeV}\, \frac{N}{6} \sqrt{\frac{6~{\rm fb}} 
    {\sigma(pp \rightarrow \phi \rightarrow \gamma\gamma)}},
\label{eq:f}\\
  \frac{2 m_D + 3 m_L}{5} &\sim 80~{\rm GeV} \sqrt{\frac{6}{N}} \sqrt{
    \frac{\sigma(pp \rightarrow \phi \rightarrow \gamma\gamma)}{6~{\rm fb}}},
\label{eq:m-1}
\end{align}
where we have used Eq.~(\ref{eq:scales}) in the second equation.  The 
ratios of branching fractions to various $\phi$ decay modes are given by
\begin{gather}
\begin{aligned}
  & \frac{B_{\phi \rightarrow gg}}{B_{\phi \rightarrow \gamma\gamma}} 
  = 8 \left( \frac{6 g_3^2}{14 e^2} \right)^2 
  \simeq 200,
\qquad
 && \frac{B_{\phi \rightarrow WW}}{B_{\phi \rightarrow \gamma\gamma}} 
  = 2 \left( \frac{9}{14 \sin^2\!\theta_W} \right)^2 
  \simeq 15,
\\
  & \frac{B_{\phi \rightarrow ZZ}}{B_{\phi \rightarrow \gamma\gamma}} 
  = \left( \frac{9 + 5 \tan^4\!\theta_W}{14 \tan^2\!\theta_W} \right)^2 
  \simeq 5,
\qquad
 && \frac{B_{\phi \rightarrow Z\gamma}}{B_{\phi \rightarrow \gamma\gamma}} 
  = 2 \left( \frac{9 - 5 \tan^2\!\theta_W}{14 \tan\theta_W} \right)^2 
  \simeq 2,
\label{eq:phi_ZG-GG}
\end{aligned}
\end{gather}
where $e$ and $\theta_W$ are the electromagnetic coupling and the 
Weinberg angle, respectively.

\begin{figure}[t]
\centering
  \subfigure{\includegraphics[clip,width=.49\textwidth]{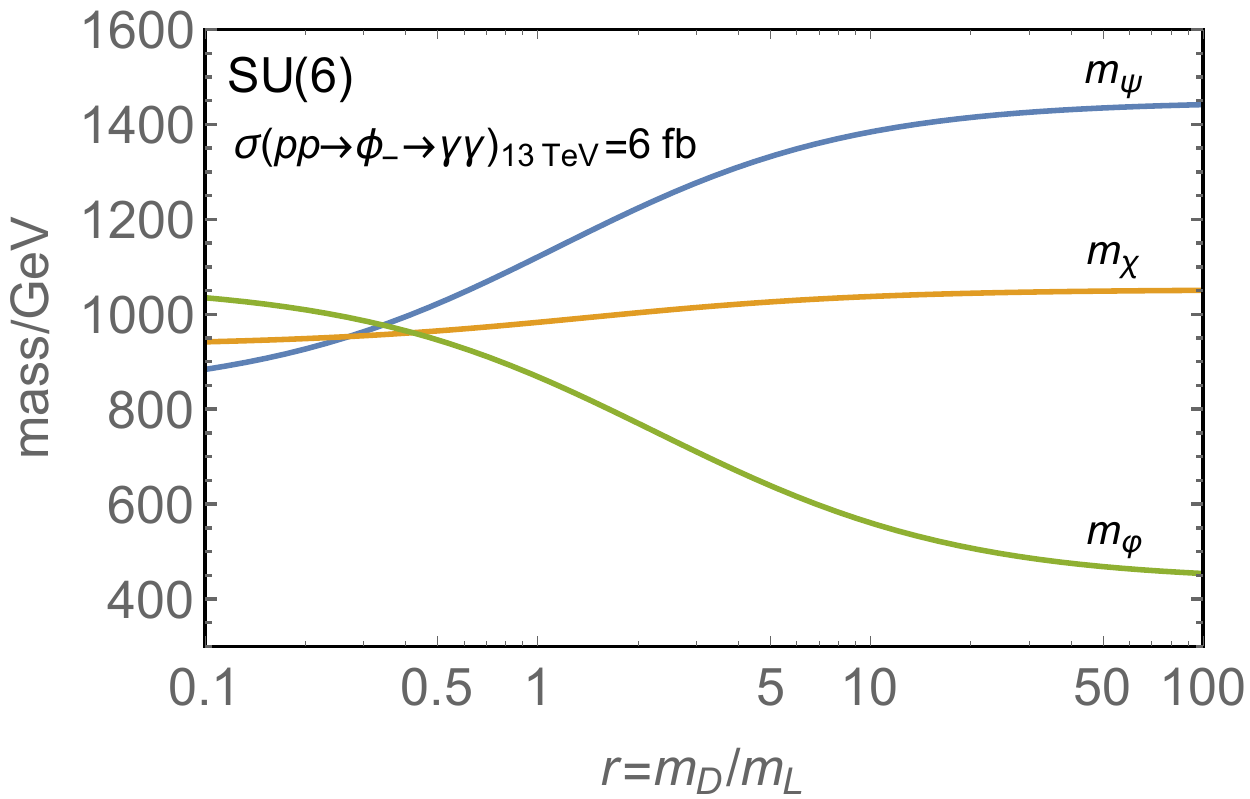}}
\hspace{3mm}
  \subfigure{\includegraphics[clip,width=.48\textwidth]{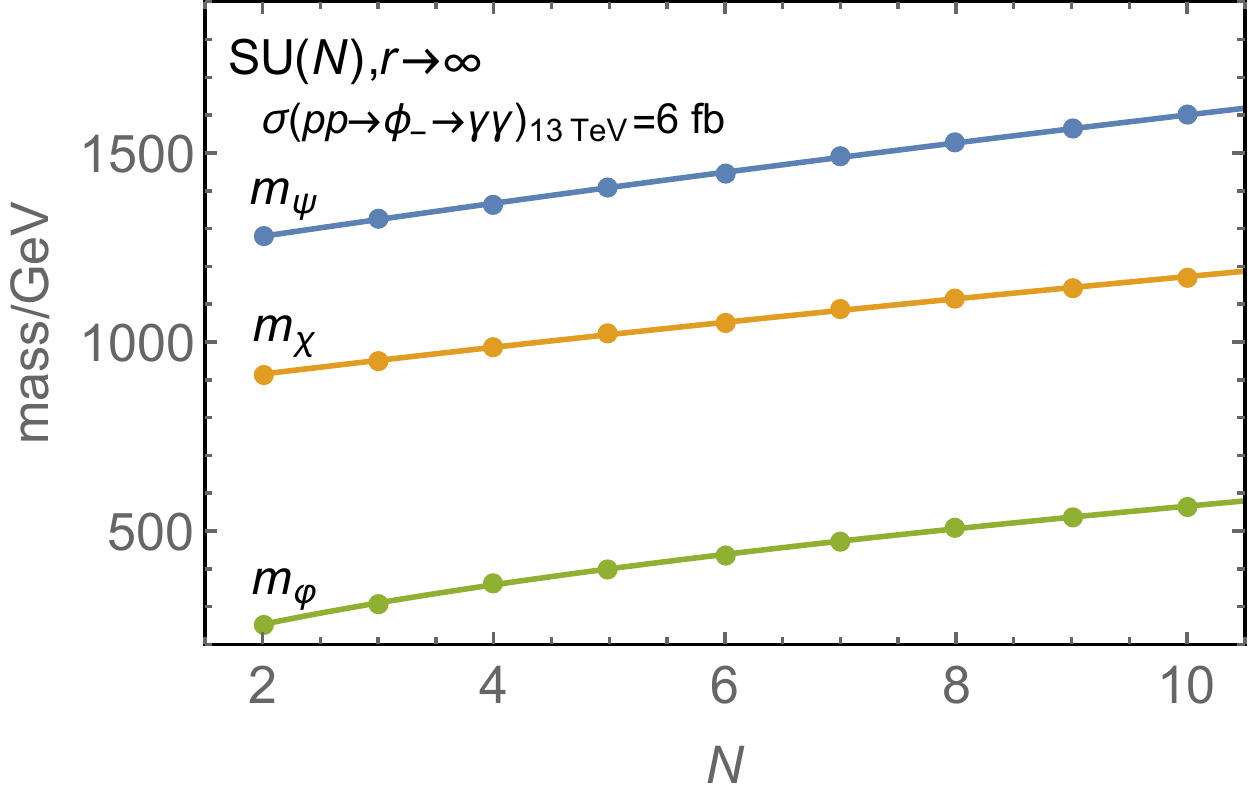}}
\caption{The masses of hidden pions $\psi$, $\chi$, and $\varphi$ 
 for $m_\phi = 750~{\rm GeV}$ as functions of $r = m_D/m_L$ for $N=6$ 
 (left) and as functions of $N$ for $r \rightarrow \infty$ (right).}
\label{fig:pion-SU}
\end{figure}
Under the conditions in Eqs.~(\ref{eq:f},~\ref{eq:m-1}), the masses 
of the other hidden pions are predicted in terms of $N$ and the ratio 
$r \equiv m_D/m_L$~\cite{Harigaya:2016pnu}.  In the left panel of 
Fig.~\ref{fig:pion-SU} we show the masses of hidden pions $\psi$, 
$\chi$, and $\varphi$ as functions of $r$ for $N = 6$.  If $m_D$ 
and $m_L$ are unified at a conventional unification scale (around 
$10^{14}\mbox{--}10^{17}~{\rm GeV}$), then their ratio at the TeV 
scale is in the range $r \simeq 1.5~\mbox{--}~3$, with the precise 
value depending on the structure of the theory above the TeV scale. 
In the right panel, we show these masses at $r \rightarrow \infty$ as 
functions of $N$.  In drawing these plots, we have taken $\Lambda = 
3.5~{\rm TeV} \sqrt{N/6}$, motivated by Eqs.~(\ref{eq:scales},~\ref{eq:f}), 
and used Eq.~(\ref{eq:Delta_CLY}) with unit coefficients.  We find 
that the colored hidden pions $\psi$ and $\chi$ are relatively light, 
$m_\psi \lesssim 1.6~{\rm TeV}$ and $m_\chi \lesssim 1.2~{\rm TeV}$, 
unless $N$ is very large, $N > 10$.  We stress, however, that these 
masses are depicted under the assumption that $\sigma(pp \rightarrow 
\phi \rightarrow \gamma \gamma)_{\rm 13\,\,TeV} = 6~{\rm fb}$.  If this 
rate is smaller, then the value of $f$, and hence $\Lambda$, becomes 
larger.  This makes the hidden pion masses larger because of larger 
gauge loop contributions.  For detailed phenomenology of these hidden 
pions, see Ref.~\cite{Harigaya:2016pnu}.

\subsection{{\boldmath $G_H = SO(N)$}}
\label{subsec:SO}

We now discuss the case with $G_H = SO(N)$.  We assume that the 
hidden quarks transform as the vector representation of $SO(N)$; 
see Table~\ref{tab:SO}.  The masses of the hidden quarks are given 
as in the $SU(N)$ case, Eq.~(\ref{eq:L_mass}).  We note that the 
matter contents in the $SO(N)$ model discussed here and the $Sp(N)$ 
model discussed in the next subsection fit into representations of 
the $SO(10)$ grand unified group, while this is not the case for 
the $SU(N)$ model.
\begin{table}[h]
\begin{center}
\begin{tabular}{c|cccc}
   & $G_H = SO(N)$ & $SU(3)_C$ & $SU(2)_L$ & $U(1)_Y$ \\ \hline
 $\Psi_D$       &       $\Box$ &  ${\bf 3}^*$ & ${\bf 1}$ &  $1/3$ \\
 $\Psi_L$       &       $\Box$ &    ${\bf 1}$ & ${\bf 2}$ & $-1/2$ \\
 $\bar{\Psi}_D$ &       $\Box$ &    ${\bf 3}$ & ${\bf 1}$ & $-1/3$ \\
 $\bar{\Psi}_L$ &       $\Box$ &    ${\bf 1}$ & ${\bf 2}$ &  $1/2$ 
\end{tabular}
\end{center}
\caption{Charge assignment of the $G_H = SO(N)$ model.  $\Psi_{D,L}$ 
 and $\bar{\Psi}_{D,L}$ are left-handed Weyl spinors.}
\label{tab:SO}
\end{table}

The approximate flavor symmetry of the $G_H$ sector is now $SU(10)$. 
This symmetry is broken to an $SO(10)$ subgroup by the hidden quark 
condensations, which can again be written as Eq.~(\ref{eq:PsiPsi-cond}) 
and do not break $G_{\rm SM}$.  We therefore have $99 - 45 = 54$ 
hidden pions.  Specifically, in addition to $\psi$, $\chi$, 
$\varphi$, and $\phi$ in Eq.~(\ref{eq:pions_SU}), we have hidden 
pions transforming as
\begin{equation}
  ({\bf 6},{\bf 1})_{-2/3}, \qquad
  ({\bf 3},{\bf 2})_{1/6}, \qquad
  ({\bf 1},{\bf 3})_1,
\label{eq:pions_SO}
\end{equation}
under $G_{\rm SM}$ (which are all complex scalars).  The masses of 
these additional hidden pions are given by
\begin{align}
  m_{({\bf 6},{\bf 1})_{-2/3}}^2 &= 2 m_D \frac{c}{f^2} 
    + \frac{10}{3} \Delta_C + \frac{4}{15} \Delta_Y,
\label{eq:m_so1}\\
  m_{({\bf 3},{\bf 2})_{1/6}}^2  &= (m_D + m_L) \frac{c}{f^2} 
    + \frac{4}{3} \Delta_C + \frac{3}{4} \Delta_L + \frac{1}{60} \Delta_Y,
\label{eq:m_so2}\\
  m_{({\bf 1},{\bf 3})_1}^2      &= 2 m_L \frac{c}{f^2} 
    + 2 \Delta_L + \frac{3}{5} \Delta_Y,
\label{eq:m_so3}
\end{align}
while those of $\psi$, $\chi$, $\varphi$, and $\phi$ are still given 
by Eqs.~(\ref{eq:m_psi}~--~\ref{eq:m_phi}).

\begin{figure}[t]
\centering
  \subfigure{\includegraphics[clip,width=.49\textwidth]{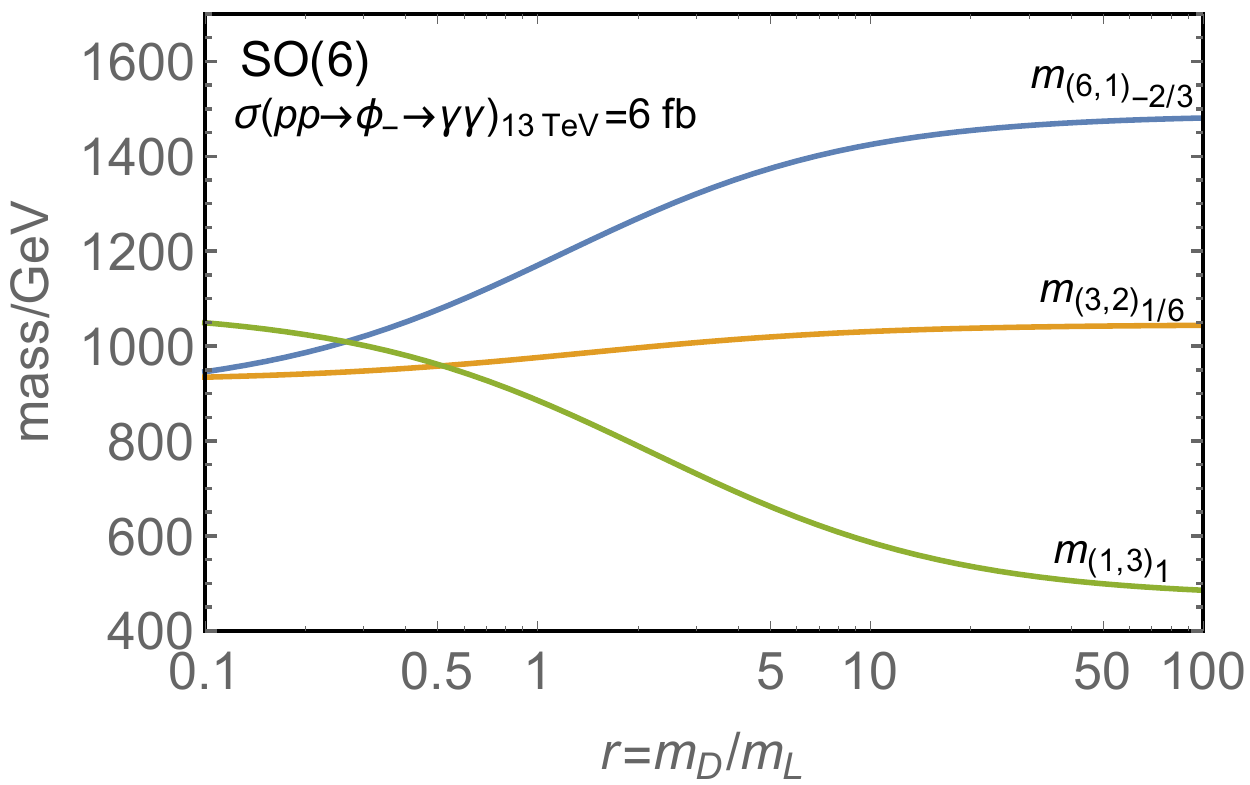}}
\hspace{3mm}
  \subfigure{\includegraphics[clip,width=.48\textwidth]{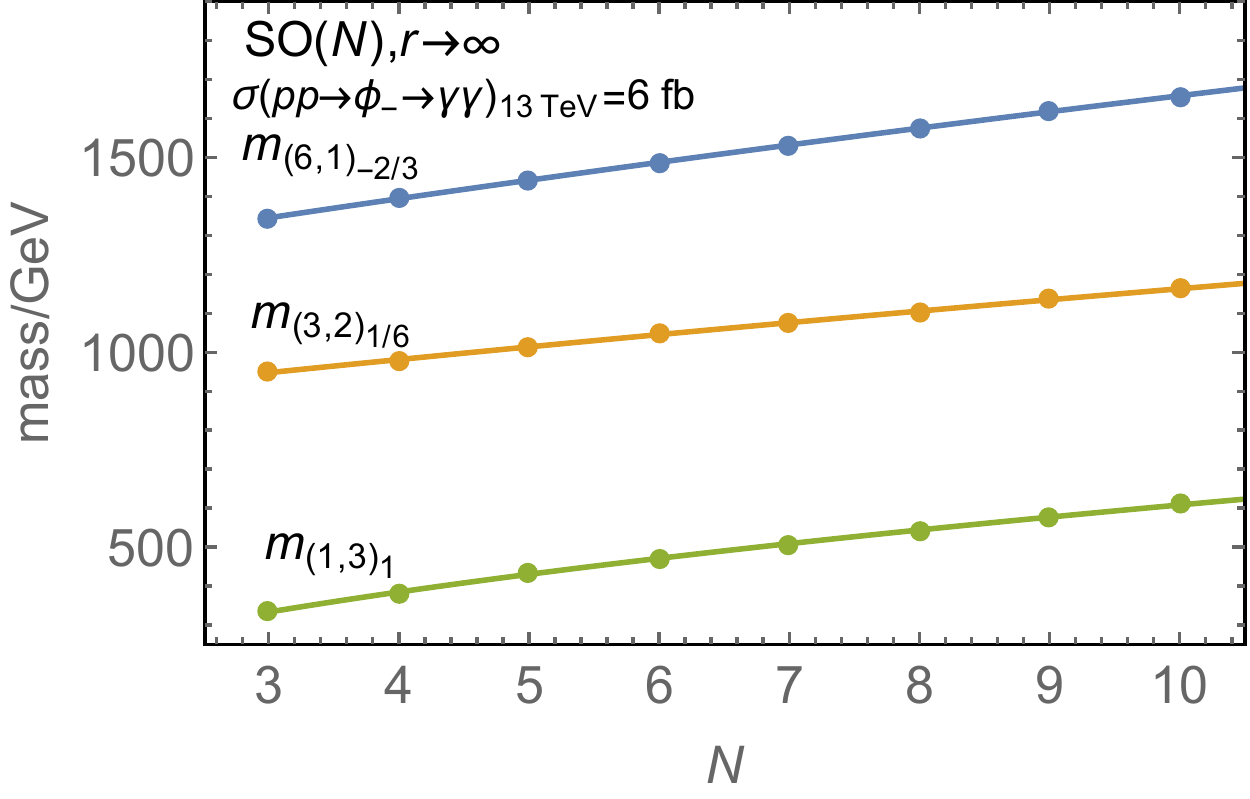}}
\caption{The masses of the $({\bf 6},{\bf 1})_{-2/3}$, 
 $({\bf 3},{\bf 2})_{1/6}$, and $({\bf 1},{\bf 3})_1$ hidden pions 
 appearing in the $SO(N)$ model, as functions of $r = m_D/m_L$ for $N=6$ 
 (left) and as functions of $N$ for $r \rightarrow \infty$ (right).}
\label{fig:pion-SO}
\end{figure}
In the left panel of Fig.~\ref{fig:pion-SO}, we show the masses of the 
additional hidden pions as functions of $r$ for $N = 6$.  In the right 
panel, we plot the masses of these hidden pions at $r \rightarrow \infty$ 
as functions of $N$.  Again, we have taken $\Lambda = 3.5~{\rm TeV} 
\sqrt{N/6}$ and used Eq.~(\ref{eq:Delta_CLY}) with unit coefficients.
We find that the additional colored hidden pions satisfy 
$m_{({\bf 6},{\bf 1})_{-2/3}} \lesssim 1.7~{\rm TeV}$ and 
$m_{({\bf 3},{\bf 2})_{1/6}} \lesssim 1.2~{\rm TeV}$ unless 
$N$ is very large, $N > 10$.  (These numbers assume $\sigma(pp 
\rightarrow \phi \rightarrow \gamma \gamma)_{\rm 13\,\,TeV} 
= 6~{\rm fb}$.)  All the additional hidden pions require extra 
interactions beyond $G_H$ gauge and standard model interactions 
to decay.  If long-lived, the $({\bf 6},{\bf 1})_{-2/3}$ and 
$({\bf 3},{\bf 2})_{1/6}$ hidden pions give phenomenology similar 
to that of long-lived $R$ hadrons in supersymmetric models, while 
the $({\bf 1},{\bf 3})_1$ hidden pion gives phenomenology similar 
to that of long-lived sleptons.

\subsection{{\boldmath $G_H = Sp(N)$}}
\label{subsec:Sp}

We finally discuss $G_H = Sp(N)$ ($N$:\ even).%
\footnote{Our notation is such that $Sp(2) \simeq SU(2)$.}
We assume that the hidden quarks transform as the fundamental 
representation of $Sp(N)$, as in Table~\ref{tab:Sp}.  The masses 
of the hidden quarks are given as in Eq.~(\ref{eq:L_mass}).%
\footnote{If we embed $G_{\rm SM}$ into the $SO(10)$ grand unified group, 
 these mass terms arise through $SO(10)$ violating effects.}
\begin{table}[h]
\begin{center}
\begin{tabular}{c|cccc}
   & $G_H = Sp(N)$ & $SU(3)_C$ & $SU(2)_L$ & $U(1)_Y$ \\ \hline
 $\Psi_D$       &       $\Box$ &  ${\bf 3}^*$ & ${\bf 1}$ &  $1/3$ \\
 $\Psi_L$       &       $\Box$ &    ${\bf 1}$ & ${\bf 2}$ & $-1/2$ \\
 $\bar{\Psi}_D$ &       $\Box$ &    ${\bf 3}$ & ${\bf 1}$ & $-1/3$ \\
 $\bar{\Psi}_L$ &       $\Box$ &    ${\bf 1}$ & ${\bf 2}$ &  $1/2$ 
\end{tabular}
\end{center}
\caption{Charge assignment of the $G_H = Sp(N)$ model.  $\Psi_{D,L}$ 
 and $\bar{\Psi}_{D,L}$ are left-handed Weyl spinors.}
\label{tab:Sp}
\end{table}

The approximate flavor symmetry of the $G_H$ sector is $SU(10)$. 
This symmetry is broken to an $Sp(10)$ subgroup by the hidden quark 
condensations, which are given by Eq.~(\ref{eq:PsiPsi-cond}) and do 
not break $G_{\rm SM}$.  We therefore have $99 - 55 = 44$ hidden 
pions.  Specifically, in addition to $\psi$, $\chi$, $\varphi$, and 
$\phi$ of the $SU(N)$ case, we have hidden pions transforming as
\begin{equation}
  ({\bf 3},{\bf 1})_{2/3}, \qquad
  ({\bf 3},{\bf 2})_{1/6}, \qquad
  ({\bf 1},{\bf 1})_1,
\label{eq:pions_Sp}
\end{equation}
under $G_{\rm SM}$ (which are all complex scalars).  The masses of 
these hidden pions are given by
\begin{align}
  m_{({\bf 3},{\bf 1})_{2/3}}^2 &= 2 m_D \frac{c}{f^2} 
    + \frac{4}{3} \Delta_C + \frac{4}{15} \Delta_Y,
\label{eq:m_sp1}\\
  m_{({\bf 3},{\bf 2})_{1/6}}^2 &= (m_D + m_L) \frac{c}{f^2} 
    + \frac{4}{3} \Delta_C + \frac{3}{4} \Delta_L + \frac{1}{60} \Delta_Y,
\label{eq:m_sp2}\\
  m_{({\bf 1},{\bf 1})_1}^2     &= 2 m_L \frac{c}{f^2} 
    + \frac{3}{5} \Delta_Y,
\label{eq:m_sp3}
\end{align}
while those of $\psi$, $\chi$, $\varphi$, and $\phi$ are given by 
Eqs.~(\ref{eq:m_psi}~--~\ref{eq:m_phi}).

\begin{figure}[t]
\centering
  \subfigure{\includegraphics[clip,width=.49\textwidth]{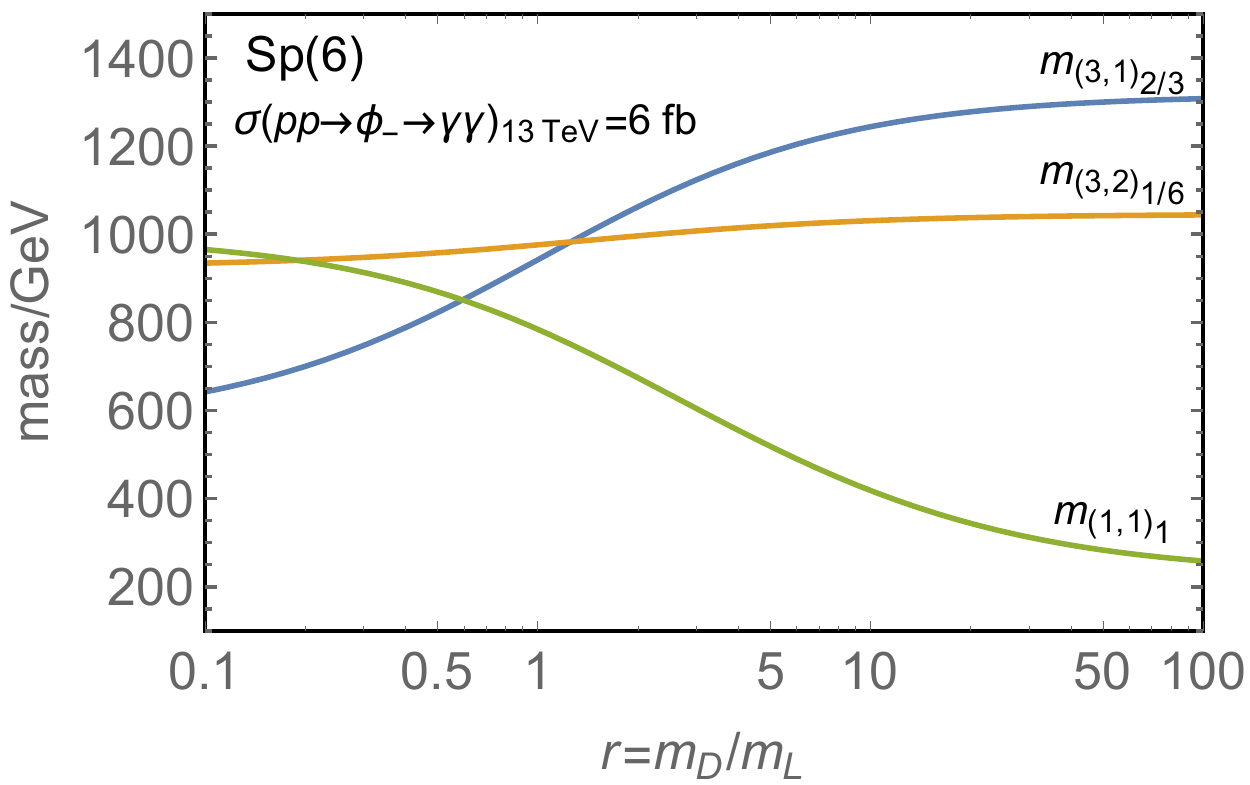}}
\hspace{3mm}
  \subfigure{\includegraphics[clip,width=.48\textwidth]{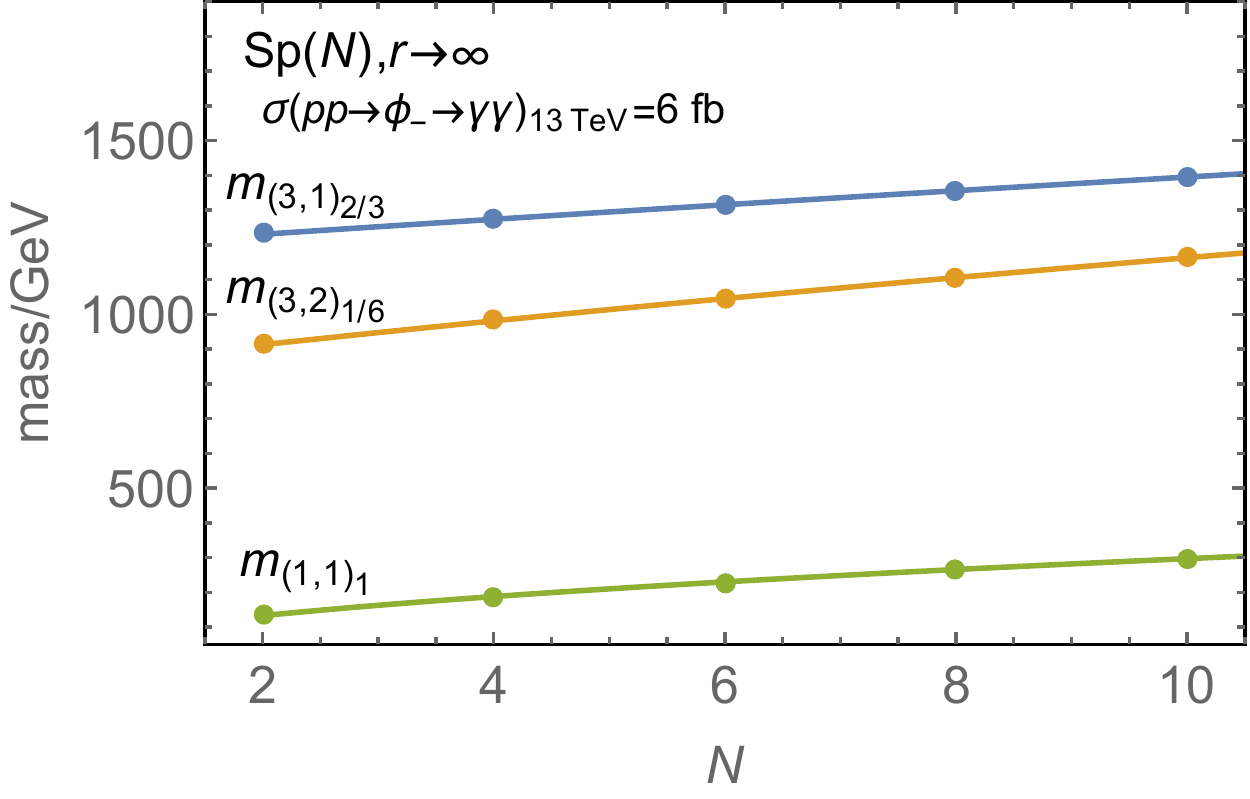}}
\caption{The masses of the $({\bf 3},{\bf 1})_{2/3}$, 
 $({\bf 3},{\bf 2})_{1/6}$, and $({\bf 1},{\bf 1})_1$ hidden pions 
 appearing in the $Sp(N)$ model, as functions of $r = m_D/m_L$ for $N=6$ 
 (left) and as functions of $N$ for $r \rightarrow \infty$ (right).}
\label{fig:pion-Sp}
\end{figure}
In the left panel of Fig.~\ref{fig:pion-Sp}, we show the masses of the 
additional hidden pions as functions of $r$ for $N = 6$.  In the right 
panel, we plot the masses of these hidden pions at $r \rightarrow \infty$ 
as functions of $N$.  We have taken $\Lambda = 3.5~{\rm TeV} \sqrt{N/6}$ 
and used Eq.~(\ref{eq:Delta_CLY}) with unit coefficients.  We find that 
the additional colored hidden pions satisfy $m_{({\bf 3},{\bf 1})_{2/3}} 
\lesssim 1.4~{\rm TeV}$ and $m_{({\bf 3},{\bf 2})_{1/6}} \lesssim 
1.2~{\rm TeV}$ unless $N$ is very large, $N > 10$.  (The numbers are 
for $\sigma(pp \rightarrow \phi \rightarrow \gamma \gamma)_{\rm 13\,\,TeV} 
= 6~{\rm fb}$.)  Again, all the additional hidden pions require extra 
interactions beyond $G_H$ gauge and standard model interactions to decay. 
If long-lived, the $({\bf 3},{\bf 1})_{2/3}$ and $({\bf 3},{\bf 2})_{1/6}$ 
hidden pions give phenomenology similar to that of long-lived $R$ hadrons, 
while the $({\bf 1},{\bf 1})_1$ hidden pion gives phenomenology similar 
to that of long-lived sleptons.

So far, we have discussed models in which $G_H = SU(N)$, $SO(N)$, and 
$Sp(N)$, and the hidden quarks are fundamental representations of $G_H$. 
The analysis can be easily extended to other representations.  Since 
chiral symmetry breaking is expected to occur in the same way (other 
than, possibly, some special cases~\cite{Kosower:1984aw}), physics of 
the hidden pions in models with hidden quarks in complex, real, and 
pseudo-real representations are identical with those of the models with 
the hidden quarks in the fundamental representations of $G_H = SU(N)$, 
$SO(N)$, and $Sp(N)$, respectively.

\section{Models with Two Diphoton Resonances from Hidden Pions}
\label{sec:non-minimal}

Here we consider models in which the two diphoton resonances arise 
from hidden pions.  In Ref.~\cite{Harigaya:2016pnu}, it was found that 
this can occur if the model has an extra hidden quark charged under 
$G_H$ but singlet under $G_{\rm SM}$.  In that paper, the introduction 
of such an extra hidden quark was motivated by cosmology.  In 
Section~\ref{subsec:1.6_SU}, we study the $G_H = SU(N)$ model in 
detail, identifying one of the two resonances as the $750~{\rm GeV}$ 
diphoton resonance.  In particular, we demonstrate how parameters of 
the models are determined and how the other hidden pion masses are 
predicted once the two diphoton resonances are observed.  We describe 
$SO(N)$ and $Sp(N)$ variants in Sections~\ref{subsec:1.6_SO} and 
\ref{subsec:1.6_Sp}, respectively.

\subsection{{\boldmath $G_H = SU(N)$}}
\label{subsec:1.6_SU}

We first discuss the case with $G_H = SU(N)$.  The matter content of the 
model is given by Table~\ref{tab:singlet}, and the masses of the hidden 
quarks are given by
\begin{equation}
  {\cal L} = -m_D \Psi_D \bar{\Psi}_D - m_L \Psi_L \bar{\Psi}_L 
    -m_N \Psi_N \bar{\Psi}_N + {\rm h.c.},
\label{eq:SU-ex_mass}
\end{equation}
where we assume $m_{D,L,N} \lesssim \Lambda$.
\begin{table}[h]
\begin{center}
\begin{tabular}{c|cccc}
   & $G_H = SU(N)$ & $SU(3)_C$ & $SU(2)_L$ & $U(1)_Y$ \\ \hline
 $\Psi_D$       &       $\Box$ &  ${\bf 3}^*$ & ${\bf 1}$ &  $1/3$ \\
 $\Psi_L$       &       $\Box$ &    ${\bf 1}$ & ${\bf 2}$ & $-1/2$ \\
 $\Psi_N$       &       $\Box$ &    ${\bf 1}$ & ${\bf 1}$ &    $0$ \\
 $\bar{\Psi}_D$ & $\bar{\Box}$ &    ${\bf 3}$ & ${\bf 1}$ & $-1/3$ \\
 $\bar{\Psi}_L$ & $\bar{\Box}$ &    ${\bf 1}$ & ${\bf 2}$ &  $1/2$ \\
 $\bar{\Psi}_N$ & $\bar{\Box}$ &    ${\bf 1}$ & ${\bf 1}$ &    $0$ 
\end{tabular}
\end{center}
\caption{Charge assignment of the $SU(N)$ model for two diphoton 
 resonances.  $\Psi_{D,L,N}$ and $\bar{\Psi}_{D,L,N}$ are left-handed 
 Weyl spinors.}
\label{tab:singlet}
\end{table}

The spectrum below $\Lambda$ consists of hidden pions
\begin{gather}
  \psi({\bf 8}, {\bf 1})_0, \qquad
  \chi({\bf 3}, {\bf 2})_{-5/6}, \qquad
  \varphi({\bf 1}, {\bf 3})_0, \qquad
  \phi({\bf 1}, {\bf 1})_0,
\nonumber \\
  \xi({\bf 3},{\bf 1})_{-1/3}, \qquad
  \lambda({\bf 1},{\bf 2})_{1/2}, \qquad
  \eta({\bf 1}, {\bf 1})_0,
\label{eq:6F-pions}
\end{gather}
where $\chi$, $\xi$, and $\lambda$ are complex while the others are 
real.  Note that there are two hidden pions which are singlet under 
$G_{\rm SM}$:\ one charged under $SU(5) \supset G_{\rm SM}$, $\phi$, 
and the other singlet under it, $\eta$.  The masses of the hidden pions 
are given by
\begin{align}
  m_\psi^2 &= 2 m_D \frac{c}{f^2} 
    + 3 \Delta_C,
\label{eq:SU6-psi}\\
  m_\chi^2 &= (m_D + m_L) \frac{c}{f^2} 
    + \frac{4}{3} \Delta_C + \frac{3}{4} \Delta_L + \frac{5}{12} \Delta_Y,
\label{eq:SU6-chi}\\
  m_\varphi^2 &= 2 m_L \frac{c}{f^2} 
    + 2 \Delta_L,
\label{eq:SU6-varphi}\\
  m_\xi^2 &= (m_D + m_N) \frac{c}{f^2} 
    + \frac{4}{3} \Delta_C+ \frac{1}{15} \Delta_Y,
\label{eq:SU6-xi}\\
  m_\lambda^2 &= (m_L + m_N) \frac{c}{f^2} 
    + \frac{3}{4} \Delta_L+ \frac{3}{20} \Delta_Y,
\label{eq:SU6-lambda}\\
  \left( \begin{array}{cc}
    m_\phi^2 & m_{\phi\eta}^2 \\
    m_{\eta\phi}^2 & m_\eta^2 
  \end{array} \right) 
  &= \left( \begin{array}{cc}
    \frac{2}{5}(2 m_D + 3 m_L) & \frac{2}{5}(m_D - m_L) \\
    \frac{2}{5}(m_D - m_L) & \frac{1}{15}(3 m_D + 2 m_L + 25 m_N) 
  \end{array} \right) \frac{c}{f^2},
\label{eq:SU6-phi-eta}
\end{align}
where $c$ and $f$ are the hidden quark bilinear condensate and the 
decay constant, respectively.

The mixing between $\phi$ and $\eta$ vanishes for $m_D=m_L$ due 
to the enhanced $SU(5)$ flavor symmetry.  Expect for this special 
case, the mass eigenstates $\phi_+$ and $\phi_-$ are determined by 
Eq.~(\ref{eq:SU6-phi-eta}) as
\begin{equation}
  \phi_+ = \eta \cos\theta + \phi \sin\theta,
\qquad
  \phi_- = -\eta \sin\theta + \phi \cos\theta.
\label{eq:singlet_mix}
\end{equation}
Here, the mixing angle $\theta$ and the mass eigenvalues $m_+$ and $m_-$ 
are related with $m_{D,L,N}$ as
\begin{align}
  m_D &= \frac{f^2}{c} \frac{m_-^2 - 3 (m_+^2-m_-^2)\tan\theta 
    + m_+^2 \tan^2\!\theta}{2 (1 + \tan^2\!\theta)},
\label{eq:SU6-mD}\\
  m_L &= \frac{f^2}{c} \frac{m_-^2 + 2 (m_+^2-m_-^2)\tan\theta 
    + m_+^2 \tan^2\!\theta}{2 (1 + \tan^2\!\theta)},
\label{eq:SU6-mL}\\
  m_N &= \frac{f^2}{c} \frac{6 m_+^2 - m_-^2 + (m_+^2-m_-^2)\tan\theta 
    + (6 m_-^2-m_+^2) \tan^2\!\theta}{10 (1 + \tan^2\!\theta)}. 
\end{align}

\begin{figure}[t]
\centering
  \includegraphics[width=0.5\linewidth]{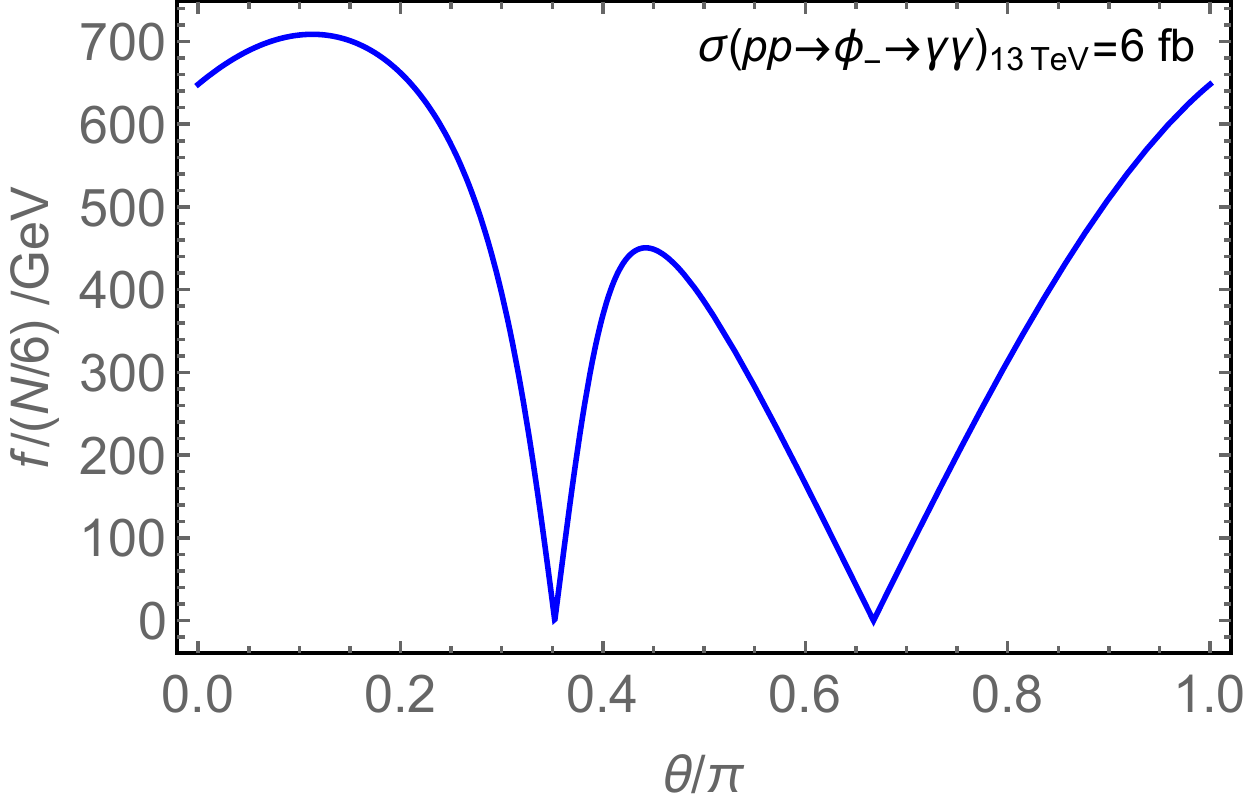}
\caption{The required value of the decay constant $f$ to obtain the 
 $750~{\rm GeV}$ diphoton rate, $\sigma(pp \rightarrow \phi_- \rightarrow 
 \gamma\gamma)_{\rm 13\,\,TeV} \simeq 6~{\rm fb}$.}
\label{fig:f_req}
\end{figure}

The dimension-five couplings of the hidden pions with the standard model 
gauge fields are determined by chiral anomalies.  The couplings of 
$\psi$, $\varphi$, and $\phi$ are given by Eq.~(\ref{eq:pion-couplings}), 
while those of $\eta$ are given by
\begin{equation}
  {\cal L} \approx \frac{N g_3^2}{64\sqrt{15}\pi^2 f} \eta\, 
    \epsilon^{\mu\nu\rho\sigma} G^a_{\mu\nu} G^a_{\rho\sigma} 
  + \frac{N g_2^2}{64\sqrt{15}\pi^2 f} \eta\, 
    \epsilon^{\mu\nu\rho\sigma} W^\alpha_{\mu\nu} W^\alpha_{\rho\sigma} 
  + \frac{N g_1^2}{64\sqrt{15}\pi^2 f} \eta\, 
    \epsilon^{\mu\nu\rho\sigma} B_{\mu\nu} B_{\rho\sigma}.
\label{eq:eta-couplings}
\end{equation}
The couplings of the mass eigenstates $\phi_{\pm}$ can be read off from 
these expressions and the mixing in Eq.~(\ref{eq:singlet_mix}).  By 
requiring that $\phi_-$ reproduces the $750~{\rm GeV}$ signal
\begin{equation}
  m_- \simeq 750~{\rm GeV},
\qquad
  \sigma(pp \rightarrow \phi_- \rightarrow \gamma\gamma)_{\rm 13\,\,TeV} 
  \simeq 6~{\rm fb},
\label{eq:phi_-}
\end{equation}
we can determine the value of $f$ as a function of the mixing angle 
$\theta$ between $\phi_\pm$.  This is plotted in Fig.~\ref{fig:f_req}.
\begin{figure}[t]
\centering
  \includegraphics[width=0.5\linewidth]{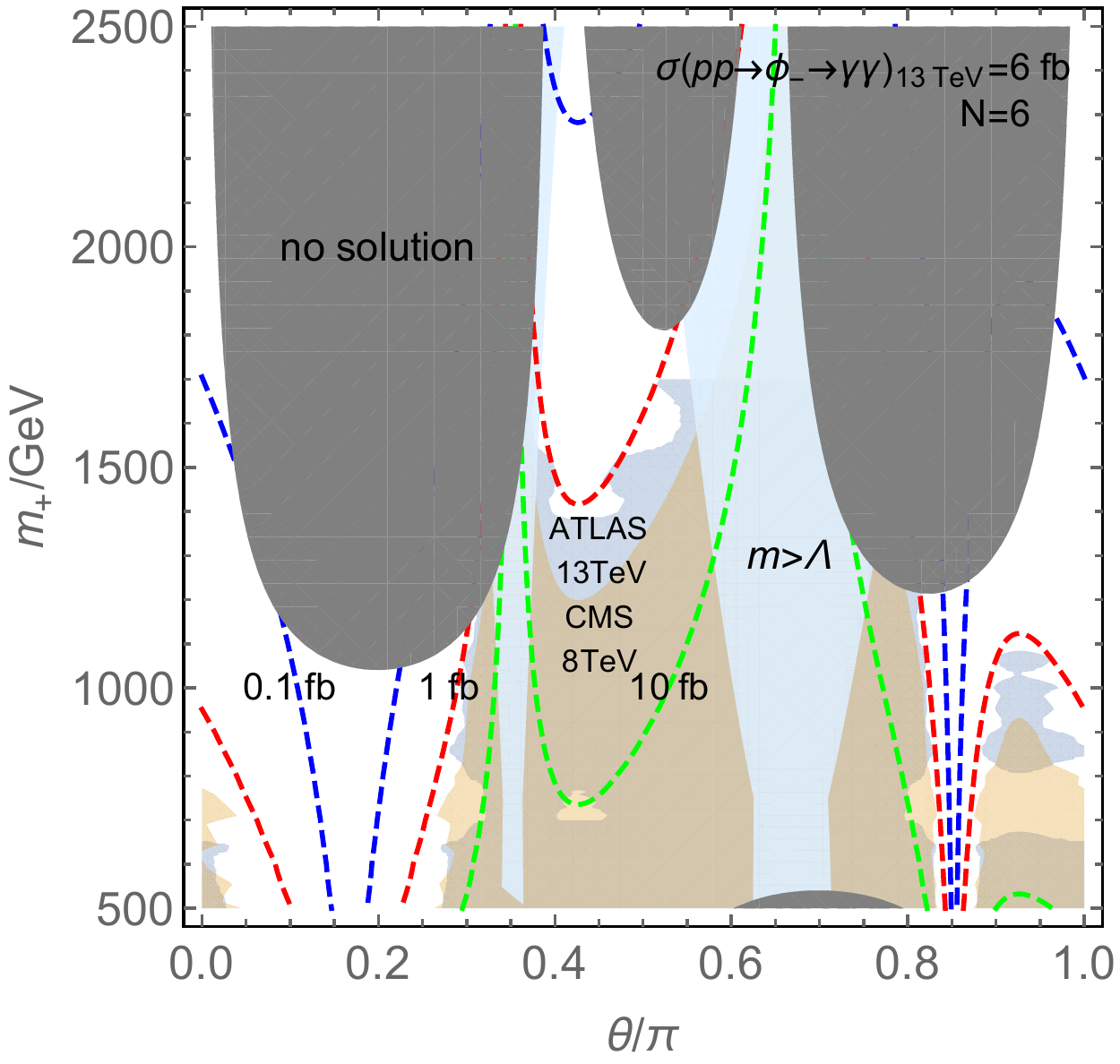}
\caption{contours of $\sigma(pp \rightarrow \phi_+ \rightarrow 
 \gamma\gamma)_{\rm 13\,\,TeV}$ in the $\theta/\pi$-$m_+$ 
 plane under the conditions that $m_- \simeq 750~{\rm GeV}$ and 
 $\sigma(pp \rightarrow \phi_- \rightarrow \gamma\gamma)_{\rm 13\,\,TeV} 
 = 6~{\rm fb}$.  The dark shaded region cannot be obtained under 
 $m_- \simeq 750~{\rm GeV}$, while the light shaded region is either 
 excluded by the ATLAS $13~{\rm TeV}$ or CMS $8~{\rm TeV}$ data or 
 outside the regime in which the hidden pion picture is valid.}
\label{fig:exclude}
\end{figure}
Under the conditions of Eq.~(\ref{eq:phi_-}), the properties of 
the second eigenstate $\phi_+$ are determined by $m_+$ and $\theta$. 
In Fig.~\ref{fig:exclude}, we show the contours of $\sigma(pp 
\rightarrow \phi_+ \rightarrow \gamma\gamma)_{\rm 13\,\,TeV}$ 
in the $\theta/\pi$-$m_+$ plane.  The dark shaded region cannot be 
obtained under $m_- \simeq 750~{\rm GeV}$.  The light shaded region 
is either excluded by the ATLAS $13~{\rm TeV}$~\cite{ATLAS-2} or 
CMS $8~{\rm TeV}$~\cite{CMS:2015cwa} data or outside the regime 
in which the hidden pion picture is valid.  (As mentioned in 
Ref.~\cite{Harigaya:2016pnu}, it is possible that $m_- \simeq 
m_+ \simeq 750~{\rm GeV}$, explaining the apparent wide width 
of the $750~{\rm GeV}$ excess.  Our present analysis does not 
include this case, which requires us to take $\sigma(pp \rightarrow 
\phi_- \rightarrow \gamma\gamma)_{\rm 13\,\,TeV} < 6~{\rm fb}$.)

There are two qualitatively different regions to notice.  In the first 
region
\begin{equation}
  -0.1 \,(= 0.9) \lesssim \frac{\theta}{\pi} \lesssim 0.3,
\label{eq:theta-1}
\end{equation}
$\phi_-$ is mostly $\phi$ while $\phi_+$ is mostly $\eta$.  In this 
region, we have viable parameter space for a wide range of $m_+$; 
in particular, for $0 \lesssim \theta/\pi \lesssim 0.3$, $m_+$ 
can be smaller than a TeV.  The diphoton rate $\sigma(pp \rightarrow 
\phi_+ \rightarrow \gamma\gamma)_{\rm 13\,\,TeV}$ is of 
$O(0.1~\mbox{--}~1~{\rm fb})$.  In the second region
\begin{equation}
  0.4 \lesssim \frac{\theta}{\pi} \lesssim 0.5,
\label{eq:theta-2}
\end{equation}
the mass of $\phi_+$ must be large, $m_+ \gtrsim 1.4~{\rm TeV}$.  In 
this region, the identity of the two eigenstates $\phi_\pm$ is almost 
opposite to the case above:\ $\phi_- \sim \eta$ and $\phi_+ \sim \phi$. 
The diphoton rate $\sigma(pp \rightarrow \phi_+ \rightarrow 
\gamma\gamma)_{\rm 13\,\,TeV}$ is again of $O(0.1~\mbox{--}~1~{\rm fb})$.

\begin{figure}[t]
\centering
  \includegraphics[width=0.51\linewidth]{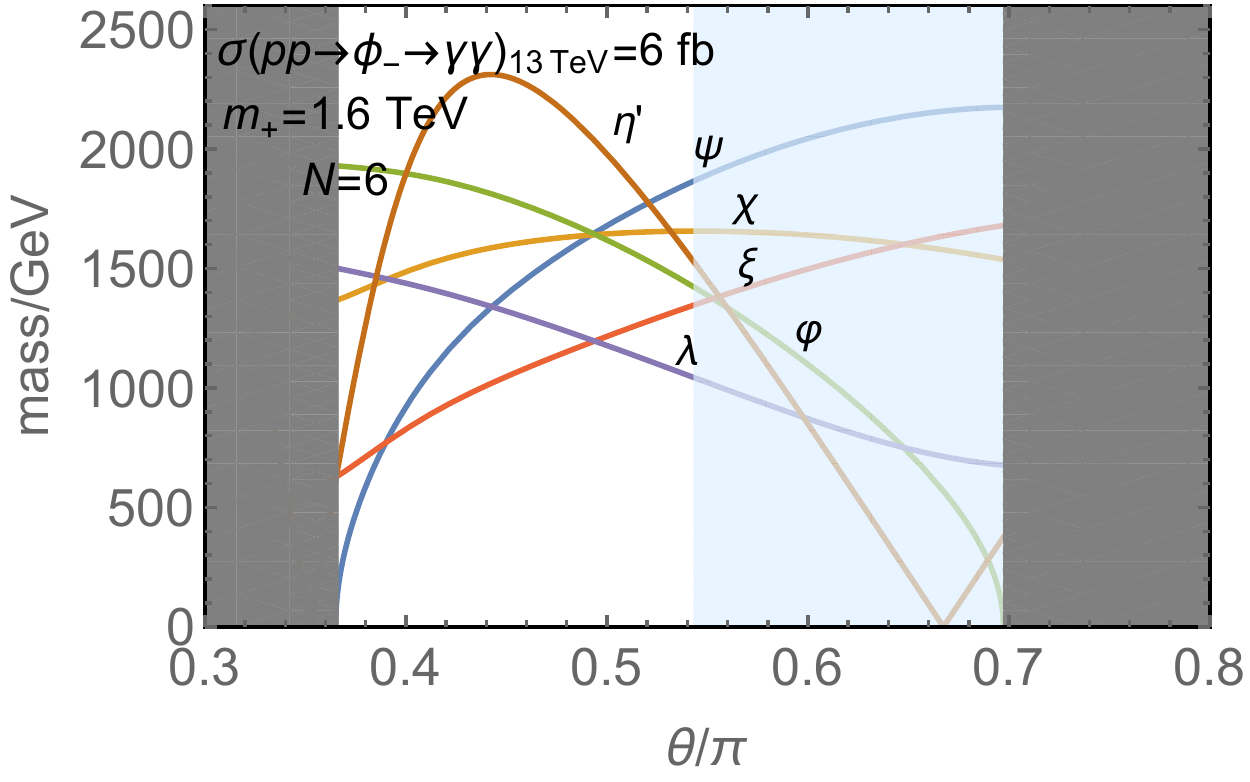}
\caption{The masses of hidden pions $\psi$, $\chi$, $\varphi$, $\xi$, 
 and $\lambda$ as well as the hidden $\eta'$ as functions of $\theta$. 
 Here, we have chosen $N=6$.}
\label{fig:mpi_theta}
\end{figure}

Below, we demonstrate if the second diphoton excess is observed, how 
parameters of the models are determined and how we can make further 
predictions.  For this purpose, we choose a benchmark point from the 
second region, where the diphoton rate is larger than the first region 
for the same value of $m_+$.  Motivated by a slight excess in the 
ATLAS data (although it is not significant)~\cite{ATLAS,ATLAS-2}, 
we choose
\begin{equation}
  m_+ \simeq 1.6~{\rm TeV},
\label{eq:m_+}
\end{equation}
and $\theta \sim \pi/2$.
\begin{figure}[t]
\centering
  \subfigure{\includegraphics[clip,width=.5\textwidth]{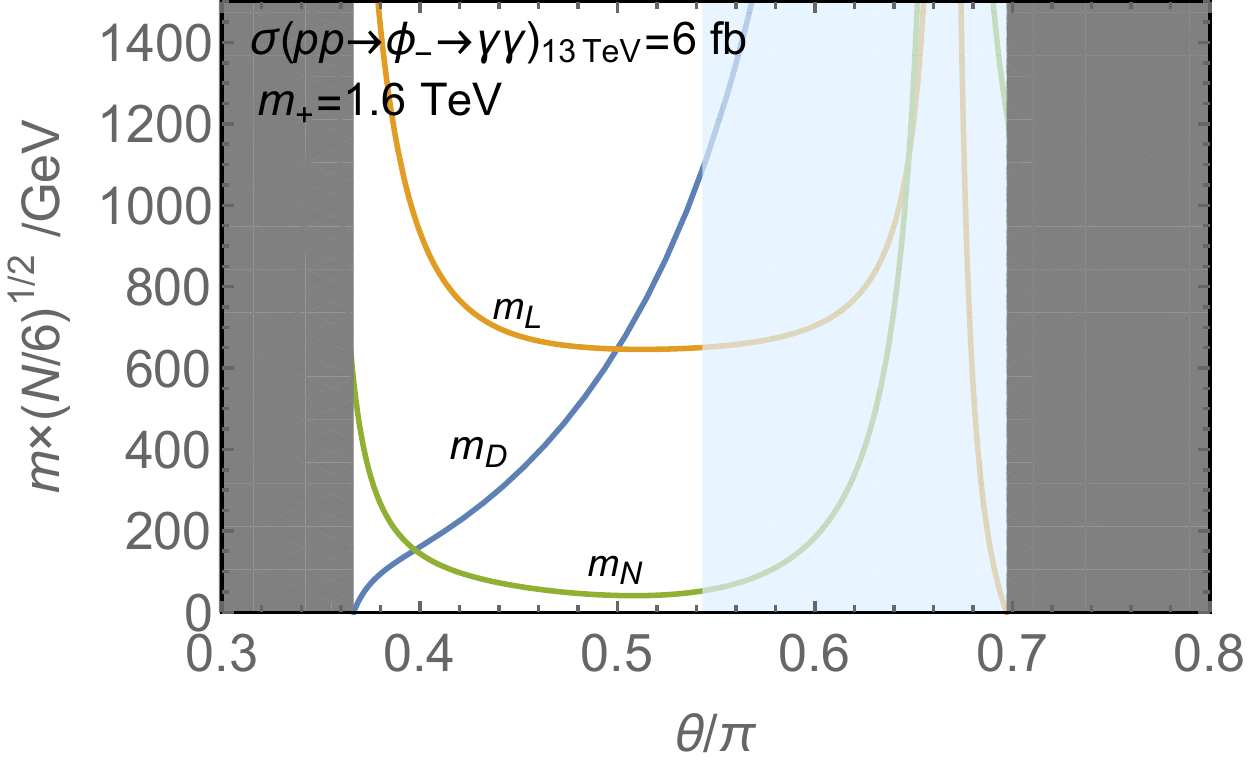}}
\hspace{3mm}
  \subfigure{\includegraphics[clip,width=.47\textwidth]{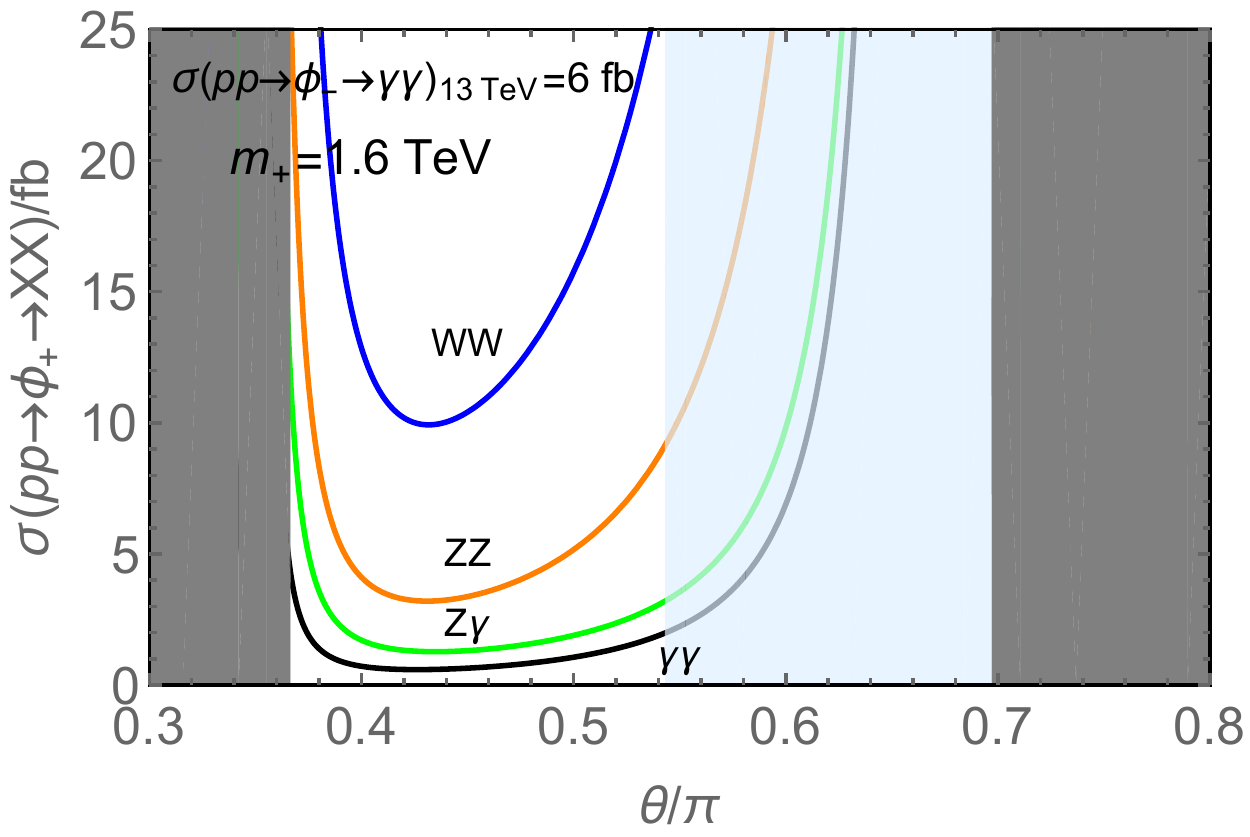}}
\caption{The hidden quark masses, $m_{D,L,N}$, that reproduce 
 $m_- = 750~{\rm GeV}$ and $m_+ = 1.6~{\rm TeV}$ as functions 
 of $\theta$ (left).  The production cross section of $\phi_+$ 
 times branching ratios into two electroweak gauge bosons at 
 $\sqrt{s} = 13~{\rm TeV}$ (right).}
\label{fig:mq-rate_theta}
\end{figure}
In the left panel of Fig.~\ref{fig:mq-rate_theta}, we plot the 
values of the hidden quark masses $m_{D,L,N}$ that reproduce 
Eqs.~(\ref{eq:phi_-},~\ref{eq:m_+}) as functions of $\theta$, 
around $\theta \sim \pi/2$.  Here, the value of the decay constant 
$f$ is determined so that $\sigma(pp \rightarrow \phi_- \rightarrow 
\gamma\gamma) = 6~{\rm fb}$ is obtained at $\sqrt{s} = 13~{\rm TeV}$; 
see Fig.~\ref{fig:f_req}.  In the dark shaded regions, no choice of 
$m_{D,L,N}$ may reproduce the required $\phi_\pm$ masses.  The light 
shaded region is excluded because of too large diphoton rates.  In 
the right panel of Fig.~\ref{fig:mq-rate_theta}, we plot predictions 
for the production cross section of $\phi_+$ times the branching 
ratios into two electroweak gauge bosons at $\sqrt{s} = 13~{\rm TeV}$. 
We find that the diphoton rate is indeed observable size, as anticipated 
from Fig.~\ref{fig:exclude}.  With Eqs.~(\ref{eq:phi_-},~\ref{eq:m_+}), 
the masses of all the other hidden pions are determined as functions 
of $\theta$, as in Fig.~\ref{fig:mpi_theta}.  Similar analyses can 
be performed for any other values of $m_+$ once the second excess 
is seen.

In Fig.~\ref{fig:mpi_theta}, we have also depicted the mass of the 
hidden $\eta'$, determined by the naive scaling~\cite{Witten:1979vv}
\begin{equation}
  m_{\eta'} = a_1 \sqrt{\frac{6}{N}}\, \Lambda,
\label{eq:m_eta'}
\end{equation}
where $a_1$ is an $O(1)$ coefficient, and $\Lambda$ is determined from 
$f$ using Eq.~(\ref{eq:scales}).  (The hidden $\eta'$ mass also has 
a contribution from the hidden quark masses, which we absorb into 
the definition of $a_1$.)  In the plot, we have chosen $a_1 = 1$ 
for definiteness.  We find that the hidden $\eta'$ is relatively light; 
this is because for $\theta \sim \pi/2$ the value of the decay constant 
$f$ required to reproduce the $750~{\rm GeV}$ excess is smaller than 
that for $\theta \sim 0$ (or in the model without an extra hidden 
quark); see Fig.~\ref{fig:f_req} and Eq.~(\ref{eq:f}).  The couplings 
of the hidden $\eta'$ with the standard model gauge bosons can be 
estimated by the $U(1)_A$ anomaly as
\begin{equation}
  {\cal L} \approx \frac{N g_3^2}{64\sqrt{3}\pi^2 f} \eta'\, 
    \epsilon^{\mu\nu\rho\sigma} G^a_{\mu\nu} G^a_{\rho\sigma} 
  + \frac{N g_2^2}{64\sqrt{3}\pi^2 f} \eta'\, 
    \epsilon^{\mu\nu\rho\sigma} W^\alpha_{\mu\nu} W^\alpha_{\rho\sigma} 
  + \frac{N g_1^2}{64\sqrt{3}\pi^2 f} \eta'\, 
    \epsilon^{\mu\nu\rho\sigma} B_{\mu\nu} B_{\rho\sigma}.
\label{eq:eta-couplings2}
\end{equation}
(This expression is valid in the large $N$ limit, and we expect that it 
gives a good approximation even for moderately large $N$.)  In particular, 
the hidden $\eta'$ also decays to a diphoton final state, with the rate 
$\sigma(pp \rightarrow \eta' \rightarrow \gamma\gamma)_{\rm 13\,\,TeV} 
\approx O({\rm fb})$.  While the estimate of the hidden $\eta'$ mass 
is subject to relatively large uncertainties, we expect from the plot 
that $m_{\eta'}$ is larger than $m_+$, at least, for some region of 
$\theta$ and $m_+$.  In this case, the diphoton excess associated with 
$\eta'$ appears above the second diphoton excess.

The relatively small value of $f$ also implies that the dynamical scale 
$\Lambda$ and hence the masses of higher resonances are also smaller. 
In particular, $C$-odd and $P$-odd spin-one resonances, which we call 
hidden rho mesons, have masses about $\Lambda$ and are expected to be 
lighter than in the model without an extra hidden quark.  The hidden 
rho mesons that have the same $G_{\rm SM}$ quantum numbers as the 
standard model gauge bosons mix with them and are singly produced 
at the LHC.  This gives lower bounds on the hidden rho meson masses. 
A particularly strong bound comes from the hidden rho meson that 
has the same $G_{\rm SM}$ charges as $\varphi$, which we refer to 
as $\rho_\varphi$.  This particle decays into an electroweak gauge 
boson and $\varphi$, with $\varphi$ subsequently decaying into a pair 
of electroweak gauge bosons.  (The decay of $\rho_\varphi$ into a pair 
of $\varphi$ is kinematically forbidden.)  The search for a resonance 
decaying into $W^\pm \gamma$~\cite{Aad:2014fha} excludes the 
$\rho_\varphi$ mass smaller than about $2~{\rm TeV}$.  While 
this constrains the parameter space, the model is still viable 
given the theoretical uncertainties associated with the estimate 
of the hidden rho meson masses and the fact that these masses 
are larger for larger $N$, scaling as $\sqrt{N}$ for the fixed 
diphoton phenomenology.

We finally mention that even if $CP$ is not preserved in the hidden sector, 
the decay of $\phi_+$ into two $\phi_-$'s is prohibited in the limit 
$m_D = m_L$ because of the enhanced flavor symmetry.  This limit occurs 
at $\theta/\pi \simeq 0.5$ for $m_+ = 1.6~{\rm TeV}$; see the left 
panel of Fig.~\ref{fig:mq-rate_theta}.  Therefore, for such values of 
$\theta/\pi$, the model provides the second diphoton excess from hidden 
pions even if $CP$ invariance in the hidden sector is not postulated.

\subsection{{\boldmath $G_H = SO(N)$}}
\label{subsec:1.6_SO}

For $G_H = SO(N)$, we may add a Majorana fermion $\Psi$ as an extra 
hidden quark, transforming as the vector representation of $SO(N)$.  Here we 
consider this minimal setup, shown in Table~\ref{tab:singlet-SO}, although 
it is straightforward to extend the analysis in the case of multiple 
extra hidden quarks.
\begin{table}[h]
\begin{center}
\begin{tabular}{c|cccc}
   & $G_H = SO(N)$ & $SU(3)_C$ & $SU(2)_L$ & $U(1)_Y$ \\ \hline
 $\Psi_D$       &       $\Box$ &  ${\bf 3}^*$ & ${\bf 1}$ &  $1/3$ \\
 $\Psi_L$       &       $\Box$ &    ${\bf 1}$ & ${\bf 2}$ & $-1/2$ \\
 $\bar{\Psi}_D$ &       $\Box$ &    ${\bf 3}$ & ${\bf 1}$ & $-1/3$ \\
 $\bar{\Psi}_L$ &       $\Box$ &    ${\bf 1}$ & ${\bf 2}$ &  $1/2$ \\
 $\Psi$         &       $\Box$ &    ${\bf 1}$ & ${\bf 1}$ &    $0$ 
\end{tabular}
\end{center}
\caption{Charge assignment of the $SO(N)$ model for two diphoton 
 resonances.  $\Psi_{D,L}$, $\bar{\Psi}_{D,L}$, and $\Psi$ are 
 left-handed Weyl spinors.}
\label{tab:singlet-SO}
\end{table}

The masses of the hidden quarks are given by
\begin{equation}
  {\cal L} = -m_D \Psi_D \bar{\Psi}_D - m_L \Psi_L \bar{\Psi}_L 
    - \frac{m}{2} \Psi^2 + {\rm h.c.},
\label{eq:SO-ex_mass}
\end{equation}
where we assume $m_{D,L}, m \lesssim \Lambda$.  The hidden quark 
condensations 
\begin{equation}
  \langle \Psi_D \bar{\Psi}_D + \Psi_D^\dagger \bar{\Psi}_D^\dagger \rangle 
  \approx \langle \Psi_L \bar{\Psi}_L 
    + \Psi_L^\dagger \bar{\Psi}_L^\dagger \rangle 
  \approx \frac{1}{2} \langle \Psi^2 + \Psi^{\dagger 2} \rangle 
  \equiv - c,
\label{eq:PsiPsi-SO-ex}
\end{equation}
breaks the approximate $SU(11)$ flavor symmetry to $SO(11)$, so we 
have $120 - 55 = 65$ hidden pions.  Specifically, in addition to 
$\psi$, $\chi$, $\varphi$, $\phi$, $\xi$, $\lambda$, and $\eta$ of 
the $SU(N)$ case (see Eq.~(\ref{eq:6F-pions})), we have hidden pions 
in Eq.~(\ref{eq:pions_SO}).

\begin{figure}[t]
\centering
  \includegraphics[width=0.51\linewidth]{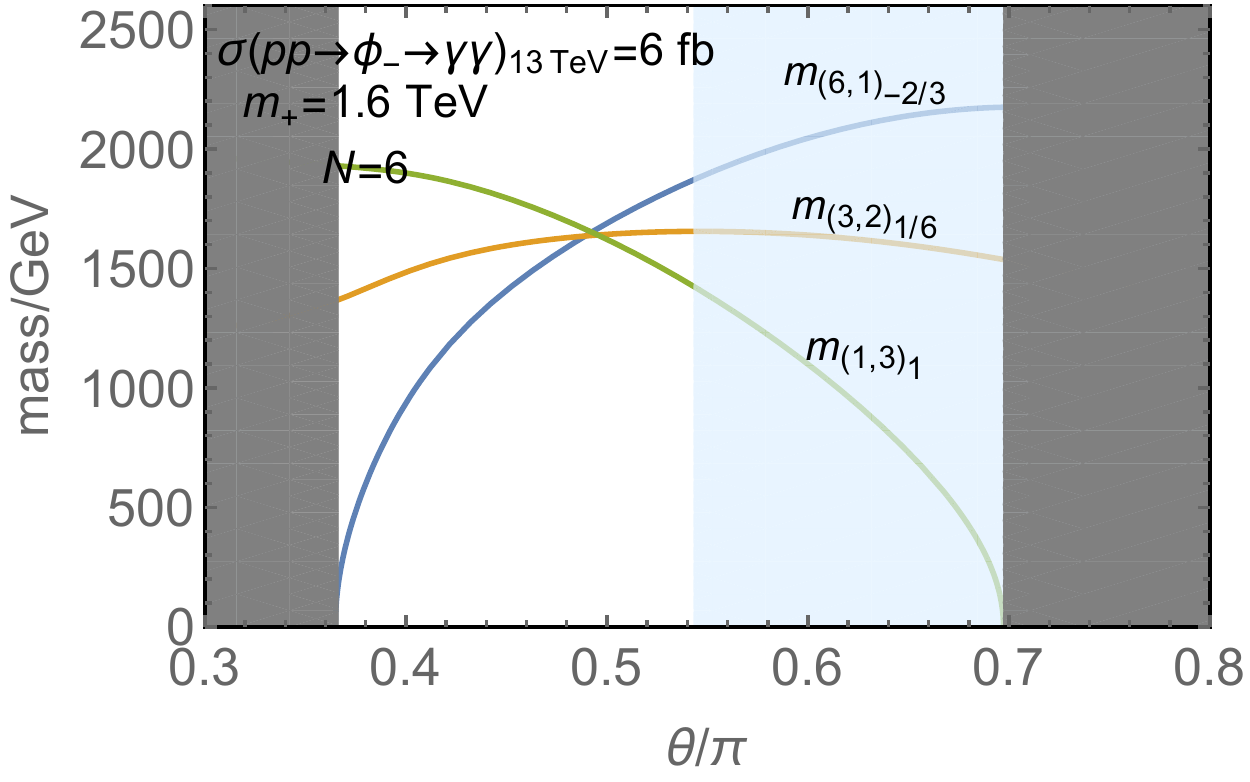}
\caption{The masses of the $({\bf 6},{\bf 1})_{-2/3}$, 
 $({\bf 3},{\bf 2})_{1/6}$, and $({\bf 1},{\bf 3})_1$ hidden pions 
 appearing in the $SO(N)$ model with an extra hidden quark $\Psi$ 
 as functions of $\theta$ for $m_+ = 1.6~{\rm TeV}$.}
\label{fig:mpi_theta-SO}
\end{figure}
Physics of the diphoton resonances is essentially the same as in the 
$SU(N)$ case, so we can simply repeat the analysis in the case of 
$G_H = SU(N)$.  For example, if we set $m_+ = 1.6~{\rm TeV}$, the 
resulting hidden pion spectrum is as given in Fig.~\ref{fig:mpi_theta} 
for $\psi$, $\chi$, $\varphi$, $\phi$, $\xi$, $\lambda$, and 
$\eta$ and in Fig.~\ref{fig:mpi_theta-SO} for the $SO(N)$ pions 
of $({\bf 6},{\bf 1})_{-2/3}$, $({\bf 3},{\bf 2})_{1/6}$, and 
$({\bf 1},{\bf 3})_1$.

\subsection{{\boldmath $G_H = Sp(N)$}}
\label{subsec:1.6_Sp}

For $G_H = Sp(N)$, the condition of global anomaly 
cancellation~\cite{Witten:1982fp} requires us to introduce 
two Weyl fermions $\Psi_{1,2}$ that transform as the fundamental 
representation of $Sp(N)$.  The matter content of this minimally 
extended model is given in Table~\ref{tab:singlet-Sp}.
\begin{table}[h]
\begin{center}
\begin{tabular}{c|cccc}
   & $G_H = Sp(N)$ & $SU(3)_C$ & $SU(2)_L$ & $U(1)_Y$ \\ \hline
 $\Psi_D$       &       $\Box$ &  ${\bf 3}^*$ & ${\bf 1}$ &  $1/3$ \\
 $\Psi_L$       &       $\Box$ &    ${\bf 1}$ & ${\bf 2}$ & $-1/2$ \\
 $\bar{\Psi}_D$ &       $\Box$ &    ${\bf 3}$ & ${\bf 1}$ & $-1/3$ \\
 $\bar{\Psi}_L$ &       $\Box$ &    ${\bf 1}$ & ${\bf 2}$ &  $1/2$ \\
 $\Psi_1$       &       $\Box$ &    ${\bf 1}$ & ${\bf 1}$ &    $0$ \\
 $\Psi_2$       &       $\Box$ &    ${\bf 1}$ & ${\bf 1}$ &    $0$ 
\end{tabular}
\end{center}
\caption{Charge assignment of the $Sp(N)$ model for two diphoton 
 resonances.  $\Psi_{D,L}$, $\bar{\Psi}_{D,L}$, and $\Psi_{1,2}$ are 
 left-handed Weyl spinors.}
\label{tab:singlet-Sp}
\end{table}

\begin{figure}[t]
\centering
  \includegraphics[width=0.51\linewidth]{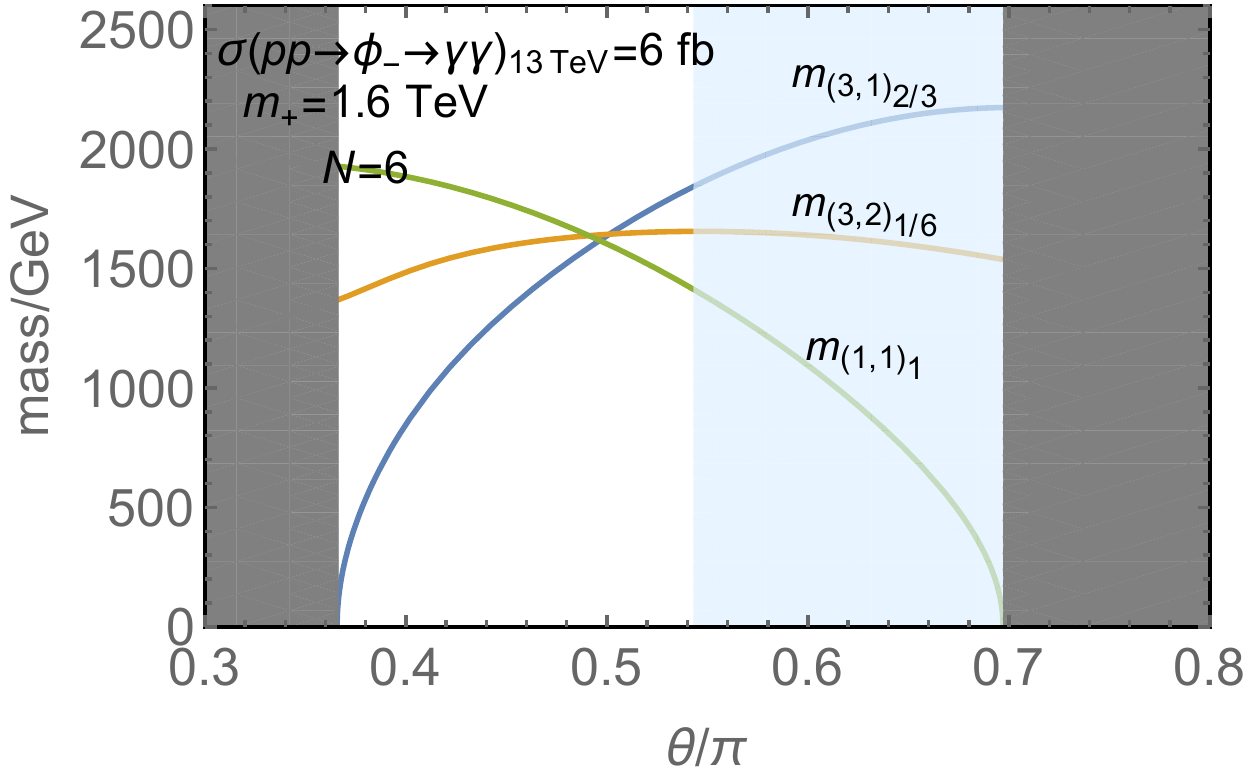}
\caption{The masses of the $({\bf 3},{\bf 1})_{2/3}$, 
 $({\bf 3},{\bf 2})_{1/6}$, and $({\bf 1},{\bf 1})_1$ hidden pions 
 appearing in the $Sp(N)$ model with extra hidden quarks $\Psi_{1,2}$ 
 as functions of $\theta$.}
\label{fig:mpi_theta-Sp}
\end{figure}

The masses of the hidden quarks are given by
\begin{equation}
  {\cal L} = -m_D \Psi_D \bar{\Psi}_D - m_L \Psi_L \bar{\Psi}_L 
    - m \Psi_1 \Psi_2 + {\rm h.c.},
\label{eq:SP-ex_mass}
\end{equation}
where we assume $m_{D,L}, m \lesssim \Lambda$.  The hidden quark 
condensations 
\begin{equation}
  \langle \Psi_D \bar{\Psi}_D + \Psi_D^\dagger \bar{\Psi}_D^\dagger \rangle 
  \approx \langle \Psi_L \bar{\Psi}_L 
    + \Psi_L^\dagger \bar{\Psi}_L^\dagger \rangle 
  \approx \langle \Psi_1 \Psi_2 + \Psi_1^\dagger \Psi_2^\dagger \rangle 
  \equiv - c,
\label{eq:PsiPsi-Sp-ex}
\end{equation}
breaks the approximate $SU(12)$ flavor symmetry to $Sp(12)$, so we 
have $143 - 78 = 65$ hidden pions.  In addition to $\psi$, $\chi$, 
$\varphi$, $\phi$, $\xi$, $\lambda$, and $\eta$ of the $SU(N)$ case, 
we now have hidden pions in Eq.~(\ref{eq:pions_Sp}) and one more set 
of $\xi$ and $\lambda$:\ $\xi'$ and $\lambda'$.  Physics of the diphoton 
resonances is again essentially the same as in the $SU(N)$ case. 
For $m_+ = 1.6~{\rm TeV}$, the masses of the $Sp(N)$ hidden 
pions, $({\bf 3},{\bf 1})_{2/3}$, $({\bf 3},{\bf 2})_{1/6}$, and 
$({\bf 1},{\bf 1})_1$, are given in Fig.~\ref{fig:mpi_theta-Sp}. 
The masses of $\xi$ and $\xi'$ and of $\lambda$ and $\lambda'$ 
are degenerate.

\section{Decays of Would-be Stable Particles}
\label{sec:decays}

The composite models we have discussed contain would-be stable particles 
which do not decay solely by $G_H$ or $G_{\rm SM}$ gauge interactions. 
Unless the reheating temperature of the universe is very small, they are 
abundantly produced in the early universe.%
\footnote{For discussions on the production of quasi-stable particles with 
 a small reheating temperature, see e.g.\ Ref.~\cite{Harigaya:2014waa}.}
If they are electrically or color charged, their lifetimes must be short 
enough to evade cosmological constraints.  In Ref.~\cite{Harigaya:2016pnu}, 
we discussed the cosmological constraints and decays of these particles 
for $G_H = SU(N)$.  Here, we extend these analyses to the case of $SO(N)$ 
and $Sp(N)$.  Unless otherwise stated, we consider models without standard 
model singlet $\Psi$'s.  We comment on the effects of the singlets 
occasionally.

We first consider hidden baryons.  In the $Sp(N)$ models, there are no 
stable baryonic composite particles.  This is a significant merit of 
the $Sp(N)$ models, since the decays of hidden baryons typically require 
operators with the dimensions higher than those of would-be stable 
hidden mesons do.

The $SO(N)$ models have baryonic states.  We assume that $m_L < m_D$ in 
the following.  For even $N$, the lightest hidden baryon is composed of 
$N/2$ $\Psi_L$ and $N/2$ $\bar{\Psi}_L$, and is neutral under the standard 
model gauge group.  Since the thermal relic abundance of this particle 
is small enough, it can be stable.  The mass difference between the 
lightest hidden baryon and other low-lying hidden baryons are of order 
$|m_D-m_L| \approx O(100~{\rm GeV})$ or $g_{1,2,3}^2 N \Lambda/16\pi^2 
\approx O(10~\mbox{--}~100~{\rm GeV})$, which is smaller than the masses 
of hidden pions.  Thus, other low-lying hidden baryons decay into the 
lightest hidden baryon and standard model particles through the emission 
of off-shell hidden pions.  Their lifetimes are as large as the lifetime 
of the emitted hidden pion.  We discuss the lifetimes of would-be stable 
pions later.  If we introduce a standard model singlet $\Psi$, then 
the lightest hidden baryon can be standard model gauge charged.  This 
is evaded if the mass of $\Psi$, $m$, is sufficiently larger than $m_L$.

For $G_H = SO(N)$ with odd $N$, the lightest baryon is composed of 
$(N+1)/2$ $\Psi_L$ and $(N-1)/2$ $\bar{\Psi}_L$ (or vice versa), which 
is charged under the standard mode gauge group as $({\bf 1},{\bf 2})_{1/2}$. 
Although the lightest baryon is electromagnetically neutral, direct 
dark matter experiments put a strict upper bound on the abundance of 
hypercharged particles as $n_B /s < 8 \times 10^{-18}$~\cite{Akerib:2015rjg}. 
The thermal abundance of the lightest hidden baryon is determined by the 
annihilation into hidden pions and is as large as $n_B/s \sim 10^{-16} 
\times (m_B / 10~{\rm TeV})$, where $m_B$ is the mass of the lightest 
hidden baryon.  The possibility of stable hidden baryons, therefore, is 
excluded.  The decay operators of hidden baryons depend of the size of 
$N$.  For $N=3$, there are four Fermi operators between three hidden 
quarks and one standard model fermion.  For $N=5$, there are operators 
composed of five hidden quarks and one standard model fermion.  If we 
introduce a standard model singlet $\Psi$ with $m \lesssim m_L$, then 
the lightest hidden baryon becomes standard model gauge group neutral, 
composed of $(N-1)/2$ $\Psi_L$, $(N-1)/2$ $\bar{\Psi}_L$, and one $\Psi$. 
In this case, the lightest hidden baryon can be cosmologically stable.

Next, we consider hidden pions.  Hidden pions which have non-zero ``$D$'' 
and/or ``$L$'' numbers are would-be stable.  For $G_H = SO(N)$ and $Sp(N)$, 
these particles can decay via the following dimension-6 operators:
\begin{equation}
  {\cal L} \sim 
    \frac{1}{M_*^{'2}} \Psi_{\bf 5} \Psi_{\bf 5} {\bf \bar{5}} {\bf \bar{5}} 
  + \frac{1}{M_*^{'2}} \Psi_{\bf \bar{5}} \Psi_{\bf \bar{5}} 
    \left({\bf \bar{5}} {\bf \bar{5}} \right)^\dag 
  + \frac{1}{M_*^{'2}} \Psi_{\bf 5}^\dag \sigma^\mu \Psi_{\bf 5} 
    {\bf \bar{5}}^\dag \sigma_\mu {\bf \bar{5}} 
  + \frac{1}{M_*^{'2}} \Psi_{\bf \bar{5}}^\dag \sigma^\mu \Psi_{\bf \bar{5}} 
    {\bf \bar{5}}^\dag \sigma_\mu {\bf \bar{5}},
\label{eq:decay-6}
\end{equation}
where we have used $SU(5)$ ($\supset G_{\rm SM}$) notation to simplify 
the expression.  If we introduce a standard model singlet $\Psi$, $L$ 
number is easily broken by a renormalizable interaction between the 
standard model Higgs, $\Psi_L$, and $\Psi$.  With this interaction, 
hidden pions that are would-be stable due to $L$ number can decay 
even promptly.

Notably, with superparticles at a low energy scale, the first term in 
Eq.~(\ref{eq:decay-6}), which leads to decays of all the would-be stable 
hidden pions, is generated from the dimension-5 superpotential term
\begin{equation}
  W = \frac{1}{M_*} \Psi_{\bf 5} \Psi_{\bf 5} {\bf \bar{5}} {\bf \bar{5}}.
\label{eq:decay-W}
\end{equation}
This should be contrasted with the case of $G_H = SU(N)$, in which writing 
down superpotential terms which have dimensions smaller than 6 and lead 
to the decay of $({\bf 3},{\bf 2})_{-5/6}$ requires an introduction of a 
standard model singlet hidden quark.  The loop of superpartners generates 
the first term in Eq.~(\ref{eq:decay-6}) with
\begin{equation}
 \frac{1}{M_*^{'2}} \sim \frac{g_H^2}{16\pi^2} \frac{1}{M_* \tilde{m}},
\label{eq:decay-scales}
\end{equation}
where $\tilde{m}$ ($> \Lambda$) represents the masses of superpartners, 
and $g_H$ is the $G_H$ gauge coupling at $\tilde{m}$.  The resultant 
lifetimes of the would-be stable hidden pions are
\begin{align}
  \tau &\sim 
    \left[ \frac{1}{8\pi} \biggl( \frac{g_H^2}{16\pi^2} 
    \frac{\sqrt{N} \Lambda^2}{4\pi M_* \tilde{m}} \biggr)^2 m_\pi \right]^{-1}
\nonumber\\
  &\sim 10^4~{\rm sec} \times 
    \biggl( \frac{M_*}{10^{16}~{\rm GeV}} \biggr)^2 
    \biggl( \frac{\tilde{m}}{10~{\rm TeV}} \biggr)^2 
    \biggl( \frac{N}{6} \biggr)^{-1} 
    \biggl( \frac{\Lambda}{3~{\rm TeV}} \biggr)^{-4} 
    \biggl( \frac{m_\pi}{1~{\rm TeV}} \bigg)^{-1} 
    \biggl( \frac{g_H}{\pi} \biggr)^{-4},
\label{eq:tau-pion}
\end{align}
where $m_\pi$ represents the hidden pion masses.  The colored 
hidden pions efficiently annihilate around the QCD phase transition 
era~\cite{Kang:2006yd}, and hence their lifetimes need only be 
smaller than $10^{13-15}~{\rm sec}$~\cite{Harigaya:2016pnu}.  This 
is easily satisfied.  On the other hand, The non-colored hidden pions 
$({\bf 1},{\bf 1})_1$ and $({\bf 1},{\bf 3})_1$ annihilate only via 
electroweak interactions.  Their abundances after freeze out are about 
$\rho/s \sim 10^{-9} ~{\rm GeV}$.  These non-colored hidden pions 
decay into a pair of lepton doublets through interactions in 
Eq.~(\ref{eq:decay-6}).  If the main decay mode is into the first 
or second generation leptons, the constraint from the big bang 
nucleosynthesis requires the lifetime to be shorter than about 
$10^4~{\rm sec}$~\cite{Kawasaki:2004qu}.  This is satisfied for 
$M_* \lesssim 10^{16}~{\rm GeV}$.  If the main decay mode is into 
the third generation leptons, then the lifetime should be shorter 
than $\approx 0.1~{\rm sec}$~\cite{Kawasaki:2004qu}, requiring smaller 
values of $M_*$.  Note that, as we have mentioned above, if we introduce 
a standard model singlet $\Psi$, then the $({\bf 1},{\bf 1})_1$ and 
$({\bf 1},{\bf 3})_1$ hidden pions can decay much more efficiently.

As discussed in Ref.~\cite{Harigaya:2016pnu}, conformal dynamics 
of $G_H$ can make $M_*$ in Eq.~(\ref{eq:decay-W}) small even if the 
suppression scale at high energies, $M_{*0}$, is large.  In fact, a 
supersymmetric $Sp(2)$ gauge theory with matter multiplets $\Psi_{D,L}$ 
and $\bar{\Psi}_{D,L}$ is in the conformal window~\cite{Seiberg:1994pq}. 
Assuming that the $Sp(2)$ gauge theory is in the conformal phase below 
the scale $M_{\rm CFT}$, $M_*$ is given by
\begin{equation}
  M_* = M_{*0} \left( \frac{\tilde{m}}{M_{\rm CFT}} \right)^{1/5} 
  = 4 \times 10^{-3} M_{*0} 
    \biggl( \frac{\tilde{m}}{10~{\rm TeV}} \biggr)^{1/5} 
    \biggl( \frac{M_{\rm CFT}}{10^{16}~{\rm GeV}} \biggr)^{-1/5}.
\label{eq:CFT}
\end{equation}
Here, we have determined the anomalous dimension of the operator 
$\Psi\Psi$ by requiring that the beta function of the $Sp(2)$ gauge 
coupling vanishes.

With conformal dynamics, dynamical scale $\Lambda$ may be related with 
the superpartner mass scale~\cite{Harigaya:2016pnu}.  The $G_H$ gauge 
theory can be in a conformal phase above $\tilde{m}$, and as the 
superpartners decouple, the theory may flow into a confining phase. 
For $G_H = Sp(N)$ with $\Psi_{D,L}$ and $\bar{\Psi}_{D,L}$, we find 
$\Lambda \sim 0.1 \tilde{m}$.

\section*{Acknowledgments}

This work was supported in part by the Department of Energy, Office of 
Science, Office of High Energy Physics, under contract No.\ DE-AC02-05CH11231, 
by the National Science Foundation under grants PHY-1316783 and PHY-1521446, 
and by MEXT KAKENHI Grant Number 15H05895.

\end{document}